\documentclass[final]{siamltex}

\usepackage{makeidx}  
\usepackage{amssymb}
\usepackage{xcolor,graphicx}
\usepackage[noend,ruled,noline,linesnumbered]{algorithm2e}

\title{Signed graph embedding: when everybody can sit closer to friends than enemies\thanks{This work has been supported by the ERC Starting research grant GOSSPLE number  204742, Comunidad de Madrid grant S2009TIC-1692, Spanish MICINN grant TIN2008--06735-C02-01 and Spanish MICINN grant Juan de la Cierva.}}

\author{Anne-Marie Kermarrec\thanks{INRIA Rennes -- Bretagne Atlantique, Campus Universitaire de Beaulieu, F-35042 Rennes, France.}
        \and Christopher Thraves\thanks{CNRS - LAAS, Univ de Toulouse - LAAS, 7 avenue du colonel Roche, F-31400 Toulouse, France.}}

\begin{document}

\maketitle

\begin{abstract}
  \emph{Signed graphs} are graphs with signed edges. They are commonly
  used to represent positive and negative relationships in social
  networks.
While \emph{balance theory}
  and \emph{clusterizable graphs} deal with signed graphs to represent social interactions, recent empirical
  studies have proved that they fail to reflect some current
  practices in real social networks. 
In this paper we address the issue of drawing signed
  graphs and capturing such social interactions. We relax the previous
  assumptions to define a drawing as a model in which every vertex has
  to be placed closer to its
  neighbors connected via a
  positive edge   
  than its 
  neighbors connected via a
  negative edge 
  in the resulting space.  Based on this definition, we
  address the problem of deciding whether a given signed graph has a
  drawing in a given $\ell$-dimensional Euclidean space. 
  We present forbidden patterns for signed graphs that admit the introduced definition of drawing in the Euclidean plane and line. 
We then focus on the $1$-dimensional case, where we provide
  a polynomial time algorithm that decides if a given \emph{complete}
  signed graph has a drawing, and constructs it when applicable. 
\end{abstract}



\pagestyle{myheadings}
\thispagestyle{plain}
\markboth{A.-~M. Kermarrec and C. Thraves}{Signed graph embedding: when everybody can sit closer to friends than enemies}

\section{Introduction}

Social interactions may reflect a wide range of relations with respect to
professional links, similar opinions, friendship, \textit{etc}. As anything
related to feelings they may well capture opposite social interactions,
e.g., like/dislike, love/hate, friend/enemy. Those social interactions
are commonly referred as binary relations. 
Recent studies on social networks have shown the existence of binary relations
%
\cite{PhysRevE.72.036121,DBLP:journals/tvcg/BrandesFL06,DBLP:conf/cse/LauterbachTSA09,DBLP:conf/chi/LeskovecHK10,DBLP:conf/icwsm/LeskovecHK10,szell10}.  
A natural
abstraction of a network that involves binary relations is a graph
with a sign assignment on their edges. Vertices related by a
positive interaction (friend) are connected via a positive edge in the
graph. On the other hand, vertices socially interacting in a negative way
(enemies) are connected via a negative edge in the graph. Such an
abstraction is known as \emph{signed graph}.

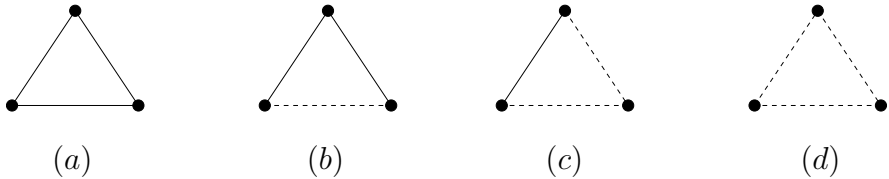
\begin{figure}[t!]
\begin{center}
\begin{picture}(0,0)(0,-10)%
\includegraphics[width=0.9\textwidth]{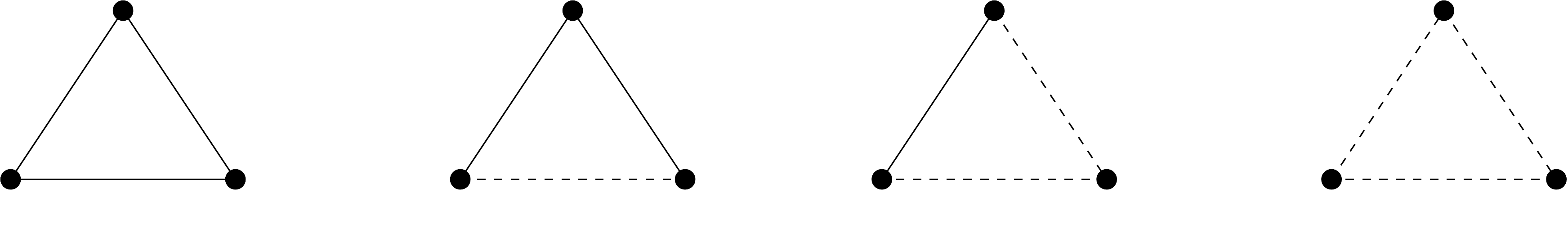}
\end{picture}%
\setlength{\unitlength}{3947sp}%
\begingroup\makeatletter\ifx\SetFigFont\undefined%
\gdef\SetFigFont#1#2#3#4#5{%
  \reset@font\fontsize{#1}{#2pt}%
  \fontfamily{#3}\fontseries{#4}\fontshape{#5}%
  \selectfont}%
\fi\endgroup%
\begin{picture}(7000,1500)(1087,-2969)
\put(4476,-2961){\makebox(0,0)[lb]{\smash{{\SetFigFont{12}{14.4}{\familydefault}{\mddefault}{\updefault}{\color[rgb]{0,0,0}$(c)$}%
}}}}
\put(2976,-2961){\makebox(0,0)[lb]{\smash{{\SetFigFont{12}{14.4}{\familydefault}{\mddefault}{\updefault}{\color[rgb]{0,0,0}$(b)$}%
}}}}
\put(1376,-2961){\makebox(0,0)[lb]{\smash{{\SetFigFont{12}{14.4}{\familydefault}{\mddefault}{\updefault}{\color[rgb]{0,0,0}$(a)$}%
}}}}
\put(6076,-2961){\makebox(0,0)[lb]{\smash{{\SetFigFont{12}{14.4}{\familydefault}{\mddefault}{\updefault}{\color[rgb]{0,0,0}$(d)$}%
}}}}
\end{picture}%
\caption{Signed triangles. This figure depicts every possible combination of positive edges and negatives edges in a triangle. 
Dashed lines represent negative edges, while continuous lines represent positive edges.}
\label{fig:triangles}
\end{center}
\end{figure}

To the best of our knowledge, the idea of signed graphs representing
social networks is introduced in the fifties by Cartwright and Harary in
\cite{Cartwright1956277}. 
In that seminal work, the notion of
\emph{balanced} signed graph is introduced and used to
characterize \textit{forbidden} patterns for social networks.
Informally, a
balanced signed graph is a signed graph that respects the following
social rules: \emph{my friend's friend is my friend, my friend's enemy
is my enemy, my enemy's friend is my enemy, and my enemy's enemy is my
friend.} Formally, a balanced signed graph is defined as a \emph{complete}
signed graph that does not contain as \emph{subgraphs}
neither triangle \textit{(b)} nor triangle \textit{(d)}
as depicted in Figure \ref{fig:triangles} (definitions for complete signed graphs and a subgraph of a signed graph are given in Section \ref{sec:problem}).  
When the signed graph is not restricted to be complete, it is said to be balanced
if all its
cycles are \emph{positive}, i.e., all its cycles have an even number of negative edges. 
Using that definition, it is
proved in \cite{harary53} that: ``A signed graph is balanced if and only if vertices can
be separated into two mutually exclusive subsets such that each
positive edge joins two vertices of the same subset and each negative
edge joins vertices from different subsets."
%

Even though, the definition of balanced signed graph intuitively makes sense to characterize social networks, one only has to consider her own set of relationships to find out that your friends are not always friends themseleves and this let us think that real life can hardly be represented by balanced structures.
Davis in \cite{davis67} gave a second characterization for social networks 
by introducing the notion of \emph{clusterizable} graph.
A signed
 graph is clusterizable if it shows a clustering, i.e., if there exists
 a partition of the vertices (may be in more than two subsets) such
 that each positive edge connects two vertices of the same subset and
 each negative edge connects vertices from different subsets.  In the
 same work, it was proved that ``a given signed graph has a clustering
 if and only if it contains no cycle having exactly one negative
 edge." Therefore, similar to the original definition of balance for complete
 signed graphs, when a given signed graph is complete it has a
 clustering if and only if it does not contain triangle \textit{(b)} in Figure
 \ref{fig:triangles} as subgraph.

Balanced signed graphs and clusterizable signed graphs have been used in
various studies about social networks \cite{DBLP:conf/chi/LeskovecHK10,DBLP:conf/icwsm/LeskovecHK10}.
Yet, the recent availability of huge databases of social networks
enabled empirical studies on real data to check if social structures
followed the balance definition
\cite{DBLP:conf/chi/LeskovecHK10,szell10}. Typically, those studies
have been carried out to check the presence or the absence of
forbidden triangles in social structures. The two common conclusions
that can be extracted from those studies are: \textit{(i)} first, the
triangle with only positive edges (triangle \textit{(a)} in
Figure \ref{fig:triangles}) is the most likely triangle to be present in
a social structure; \textit{(ii)} second, the four
possible triangles (triangle \textit{(a-d)} in Figure \ref{fig:triangles})
are always present in a social structure. Consequently, we can
arguably conclude that neither balanced signed graphs nor
clusterizable signed graphs fully represent social structures and let
us revisit the representation of social interactions.

Our contribution extends the notion of balanced and clusterizable signed
graphs. We relax the previous definition and consider that a social
structure should merely ensure that each vertex is able to have its friends
closer than its enemies.  In other words, if a signed graph is
embedded into a metric space, each vertex should be placed in the
resulting space closer to its neighbors connected via a positive edge
than to its neighbors connected via a negative edge.  
In this paper, we tackle
 this issue introducing a formal definition of a \emph{valid} graph
 drawing for signed graphs in an Euclidean metric space.  
 


The rest of the paper is structured as follows: 
After formalizing
the notion of a valid drawing for signed graphs in Section
\ref{sec:problem}, we place in perspective our definitions with respect to balanced and clusterizable signed graphs in Section \ref{subsec:balanced}.
Then, we visit the
related works in Subsection \ref{sec:relatedwork} and
we briefly describe our
contribution in Subsection \ref{subsec:contributions}. 
In Section
\ref{sec:counterexamples}, we show examples of signed graphs without a
valid drawing in the Euclidean plane, and present five infinite families of signed graphs without a valid drawing in the Euclidean line. 
We also introduce 
the notion of \emph{minimal} signed graph without a valid drawing. That notion captures the
idea of forbidden patterns in a social structure.
We then focus on dimension $1$ in Section
\ref{sec:linecase} where we fully characterize complete signed graphs that admit a valid drawing. 
%
We finally close our work in
Section \ref{sec:futurework} with a discussion about the problems this
work left open. 




%
\subsection{Problem definition}\label{sec:problem}
%

In this section we present the context of this work,  the 
required definitions, notations and  state the problem.

\paragraph{Definitions and notations}
We use $G=(V,E)$ to denote a graph where vertices are denoted with
$p_i$, and $i$ ranges from $1$ to $n$. The edge that connects vertices $p_i$ and $p_j$ is denoted $(p_i,p_j)$.  
A \emph{signed graph} is
defined as follows.
\begin{definition}\label{def:signedgraph}
A \emph{signed graph} is a graph $G=(V,E)$ together with a 
sign assignment $f: E \rightarrow \{-1,+1\}$ to its edges. 
\end{definition}

Equivalently, a signed graph can be defined as a graph $G$ together
with a bipartition of the set of edges. Using Definition
\ref{def:signedgraph}, the set of edges $E$ is partitioned in $E^+ =\{
e \in E: f(e) = +1\}$ and $E^- =\{e \in E: f(e) =-1\}$, such that
$E=E^+\bigcup E^-$ and $E^+\bigcap E^- =\emptyset$.  In the rest of
the document, we use $G=(V,E^+\bigcup E^-)$ to denote a signed graph
composed by vertex set $V$, with $E^+$ and $E^-$ as the bipartition of the edge set $E$.
%
%
%
A signed graph is \emph{complete}
if every pair of distinct vertices is connected by a unique signed edge.
The \emph{subgraph} of a given signed graph $G$ is a subgraph of $G$
that matches the sign assignment.

Given a signed graph $G=(V,E^+\bigcup E^-)$, we define
positive and negative neighbors for each vertex in $G$. Let
$p_i$ be a vertex in $G$, and let $N_i=\{p_j \in V: (p_i,p_j) \in E\}$ be
the set of neighbors of $p_i$. We define the set of \emph{positive}
and \emph{negative} neighbors of $p_i$ as follows: $N_i^+ :=\{p_j \in N_i: (p_i,p_j)
\in E^+\}$ and
  $N_i^- :=\{p_j \in N_i: (p_i,p_j)
\in E^-\}$, respectively. Therefore, vertices that belong to $N_i^+$ can be considered as $p_i$'s
friends, while 
vertices in $N_i^-$ can be considered as
$p_i$'s enemies.

A \emph{drawing} of a graph $G$ 
in the $\ell$-dimensional
Euclidean space $\mathbb{R}^\ell$ is an injection of the set of vertices $V$
in $\mathbb{R}^\ell$. 
%
Since this definition of graph drawing is not 
sufficient to capture signs in signed graphs for there is 
no  distinction between positive and negative edges,
we introduce a new specific definition of \textit{valid}
graph drawing for signed graphs.
\begin{definition}\label{def:properdrawing}
Let $G=(V,E^+\bigcup E^-)$ be a signed graph, and $D:V\rightarrow \mathbb{R}^\ell$ be a drawing of $G$ in $\mathbb{R}^\ell$. 
We say that $D(\cdot)$ is \emph{valid} 
 if and only if 
$$
d(D(p_i),D(p_j)) < d(D(p_i),D(p_k)), 
$$
for all pair of incident edges $(p_i,p_j), (p_i,p_k)$ such that  $(p_i,p_j) \in E^+$ and $(p_i,p_k) \in E^-$,
where $d(\cdot,\cdot)$ denotes the Euclidean distance between two elements in $\mathbb{R}^\ell$.  
\end{definition}

Definition \ref{def:properdrawing} captures the fact  that every vertex
has to be placed closer to its positive neighbors than to its
negative neighbors. 
In the case that there exists a valid drawing of a given signed
graph $G$ in $\mathbb{R}^\ell$, we say that $G$ \emph{has a valid
  drawing} in $\mathbb{R}^\ell$, or simply, we say that $G$ has a
valid drawing in dimension $\ell$. Otherwise, we simply say that $G$
is a signed graph without valid drawing in dimension $\ell$.
%
In this paper, we are interested 
in a \emph{classification problem}, aiming at determining if a given signed graph has a valid drawing in a given space $\mathbb{R}^\ell$. 
Particularly, we focus in the $1$-dimensional case.
 
%

\subsection{Contextualizing valid drawings with balanced and clusterizable signed graphs}
\label{subsec:balanced}
Balanced 
and clusterizable signed graphs are closely
related. Indeed, it is straightforward to observe that, 
 if we denote $\mathcal{B}$ the
set of balanced signed graphs and $\mathcal{C}$ the set of
clusterizable graphs, then $\mathcal{B}$ is a proper
subset of $\mathcal{C}$. That comes from the characterization of
balanced signed graphs as clusterizable signed graphs with at most two
clusters.

On the other hand, if we denote $\mathcal{D}^\ell$ the set of
signed graphs that have a valid drawing in the $\ell$-dimensional
Euclidean space. It is also straightforward to note that
$\mathcal{D}^\ell$ is a subset of $\mathcal{D}^{\ell'}$ if $\ell\leq
\ell'$. Hence, there is a chain of set inclusions of the form
$\mathcal{D}^1\subset \mathcal{D}^2 \subset \mathcal{D}^3 \subset
\cdots \subset \mathcal{D}^\ell \subset \cdots$.

In order to place our definition of valid drawing in the context of
balanced and clusterizable signed graphs, we point out the fact that \textit{if
a graph is clusterizable, then it has a valid drawing in the Euclidean
line}.  To demonstrate that fact, let us  draw a clusterizable
signed graph in the following way: place every vertex of a cluster
within an interval of length $d$, then every positive edge will have
length at most $d$.  Thereafter, place every pair of clusters at
distance at least $d'$  from each other, such that $d'>d$.
Therefore, every negative edge will have length at least $d'>d$. Thus, the drawing will be valid. Hence, we complete the chain of set
inclusions adding clusterizable and balanced signed graphs
as follows: $$\mathcal{B}\subset\mathcal{C}\subset \mathcal{D}^1\subset
\mathcal{D}^2 \subset \mathcal{D}^3 \subset \cdots \subset
\mathcal{D}^\ell \subset \cdots.$$ 
Establishing this connection is important and helps us
to put in context the  study of $\mathcal{D}^1$ case, 
our main contribution in this work.
\subsection{Related work}\label{sec:relatedwork}
To the best of our knowledge, the notion of balanced signed graph
is introduced by Harary in \cite{harary53}, where structural
results are presented.  Thereafter, Cartwright and Harary applied
structural balance theory to social structures, and they compared it
with Heidedr's theory in \cite{Cartwright1956277}.  Later, Davis
relaxed the definition of balanced signed graph in \cite{davis67} to
obtain clusterizable graphs, a more general structure on signed
graphs.  The aforementioned works represent the theoretical basis of
clustering and structural balance in signed graphs. On top of that, we
found several other works with interesting contributions to structural
balance theory.  Just as an example of them, in
\cite{DBLP:journals/mss/HararyK80} Harary and Kabell gave a 
polynomial time algorithm that detects balance in signed graphs,
whereas in \cite{harary81,springerlink:10.1007/BF02476926} the authors
counted balanced signed graphs using either marked graphs or P\'olya
enumeration theorem.

The clustering problem on signed graphs is studied by Bansal et al. in
\cite{DBLP:journals/ml/BansalBC04}. In that work, the authors
considered an optimization problem where the set of vertices of a
signed graph has to be partitioned such that the number of positive
edges within clusters plus the number of negative edges between
clusters is maximized.  Clusterizable signed graphs defined by Davis,
for instance, achieve the maximum of this value in the total number of
edges. Bansal et al. proved that finding the optimal clustering is
NP-hard, and they gave an approximation algorithm that minimizes
disagreements and that maximizes agreements.

To the best of our knowledge, the closest work to what we have proposed here is
\cite{DBLP:conf/sdm/KunegisSLLLA10} by Kunegis et al. In that work,
the authors applied spectral analysis to provide heuristics for
visualization of signed graphs, and link prediction. The visualization
proposed is equivalent to the visualization we propose. Nevertheless,
their work is only empirical and applied to 2D visualization, hence
our contributions complement their contributions.

Recently, signed graphs have been used to study social networks. Antal
et al. in \cite{PhysRevE.72.036121} studied the dynamic of social
networks. The authors studied the evolution of a social network
represented by a signed graph under the dynamic situation when a
link changes its sign if it is part of an imbalanced triangle.
The authors proved the
convergence to a balanced state where no imbalanced triangles
remain. Leskovec et al. in \cite{DBLP:conf/chi/LeskovecHK10} studied
how binary relations affect the structure of on-line social networks.
The author connected their analysis with structural balance theory and
other structural theories on signed graphs. One of their conclusions
says that structural balance theory captures some common patterns but
fails to do so for some other fundamental observed phenomena. 
The structure
of on-line social networks is also studied in \cite{szell10}, where
the authors study a complete, multi-relational, large social network
of a society consisting of the 300,000 players of a massive multiplayer
on-line game. The authors present an empirical verification of the
structural balance theory. 
Finally, prediction of signed links based
on balance and status theory is studied by Leskovec et al. in
\cite{DBLP:conf/www/LeskovecHK10}. The authors prove that signed
links can be predicted with high accuracy in on-line social networks
such as Epinions, Slashdot and Wikipedia.

\subsection{Our contributions}\label{subsec:contributions}

Our first contribution is precisely the definition of a 
valid drawing that
represents social structures.
As far as we know, we are the first to formally define the notion of valid drawing for
signed graphs to express the intuitive notion of \textit{everybody sits closer to friends than enemies}.
Consequently, the classification problem is considered in this setting for the first time as well. 

First, we show that some graphs do not have a valid drawing. To this
end, we provide two examples of signed graphs without valid drawing in the
Euclidean plane and four infinite families of signed graphs without valid drawing in the Euclidean line.  
Also, we introduce the concept of \emph{minimal
  signed graph without valid drawing} in a given $\ell$-dimensional
Euclidean space.  That definition helps us to find forbidden patterns
when we have to decide whether a given graph has or not a valid
drawing in a given dimension. We show that all the graphs without valid drawing presented in this work are minimal. 

   
Thereafter, we focus on the specific issue of deciding whether a
complete signed graph has a valid drawing in the Euclidean
line.  We characterize the set of complete signed graphs with a valid
drawing in the line.  We provide a polynomial time algorithm that
decides whether a given complete signed graph has or not a valid
drawing in the line. When a given complete signed graph has a valid
drawing in the line, we provide a polynomial time algorithm that
constructs a valid drawing for it. 
%

%
%
\section{Graphs without valid drawing}\label{sec:counterexamples}
In this section we present examples of signed graphs without valid drawing in the Euclidean line and plane. 
Specifically, we present two signed graphs without valid drawing in $\mathbb{R}^2$ 
and five infinite families of signed graphs without valid drawing in $\mathbb{R}$. 
We conclude this section with the introduction    
of the concept of \emph{minimal} signed graph without valid drawing, and with a discussion about its consequences.

\subsection{Signed graphs without valid drawing in the plane}
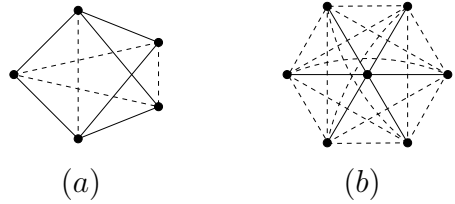
\begin{figure}[h!]
\begin{center}
\begin{picture}(0,0)(-100,0)%
\includegraphics[height=1in]{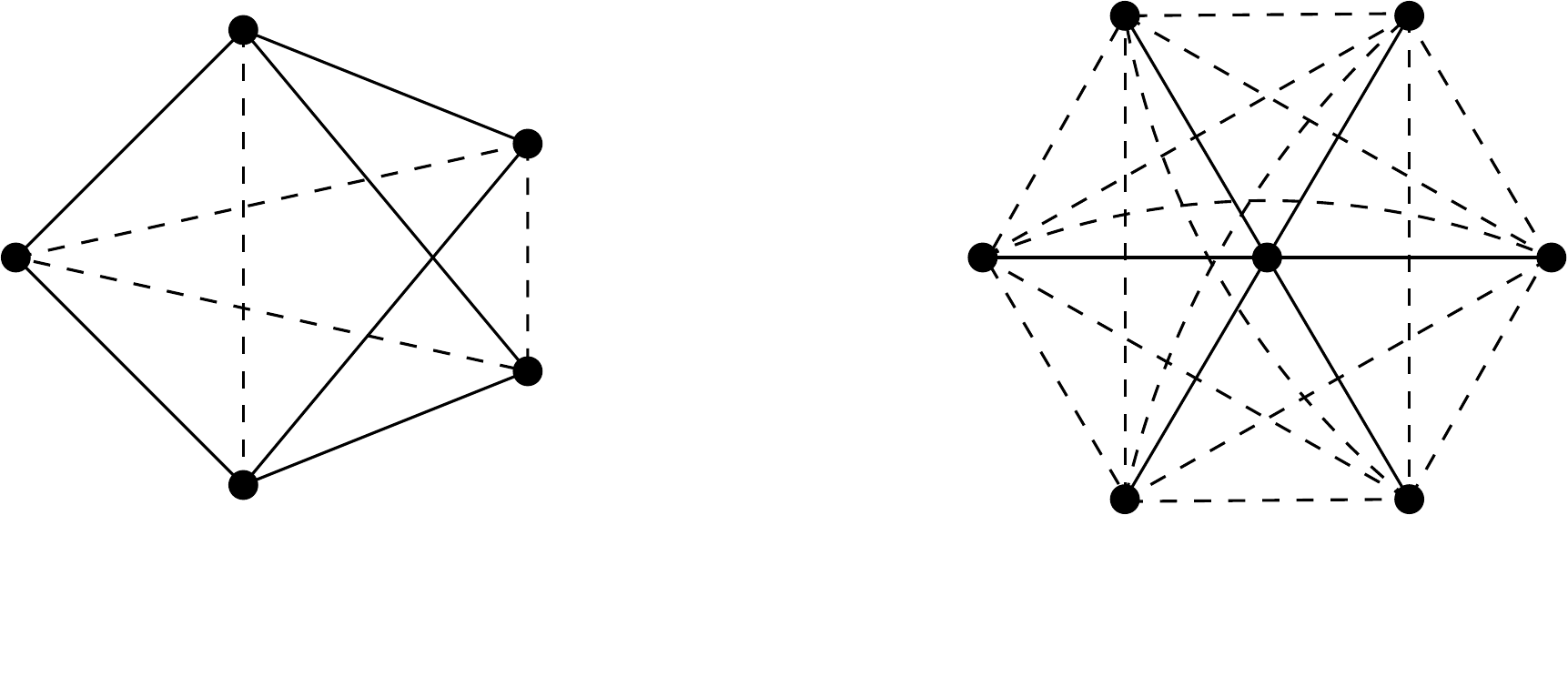}
\end{picture}%
\setlength{\unitlength}{3947sp}%
\begingroup\makeatletter\ifx\SetFigFont\undefined%
\gdef\SetFigFont#1#2#3#4#5{%
  \reset@font\fontsize{#1}{#2pt}%
  \fontfamily{#3}\fontseries{#4}\fontshape{#5}%
  \selectfont}%
\fi\endgroup%
\begin{picture}(7000,1500)(1918,-4269)
\put(5676,-4261){\makebox(0,0)[lb]{\smash{{\SetFigFont{12}{14.4}{\rmdefault}{\mddefault}{\itdefault}{\color[rgb]{0,0,0}$(b)$}%
}}}}
\put(3901,-4261){\makebox(0,0)[lb]{\smash{{\SetFigFont{12}{14.4}{\familydefault}{\mddefault}{\updefault}{\color[rgb]{0,0,0}$(a)$}%
}}}}
\end{picture}%
\caption{This figure shows \emph{negative triangle} signed graph \textit{(a)} and \emph{negative cluster} signed graph \textit{(b)}. 
 Dashed lines represent negative edges and continuous lines represent positive edges.}
\label{counterexampledim2}
\end{center}
\end{figure}

We present two signed graphs without valid drawing in the plane. We call these two signed graphs \emph{the negative triangle} and \emph{the negative cluster}. 
Graphic representations of these two signed graphs are presented in Figure \ref{counterexampledim2}. In the following, we give a detailed description of these two signed graphs and
prove why they do not have valid drawings in the plane. 
\paragraph{Negative triangle} The \emph{negative triangle} is a graph with five vertices, three out of the five vertices are connected
by a triangle of negative edges. The two remaining vertices are connected by a negative edge. The rest of the edges are positive.  
Figure \ref{counterexampledim2}\textit{(a)} depicts the negative triangle. 
\begin{lemma}\label{lem:signedhexagon} 
The negative triangle is a signed graph without valid drawing in $\mathbb{R}^2$.
\end{lemma}
\begin{proof}
The proof is by contradiction. Let us assume that there exists a valid drawing $D(G)$ in the Euclidean plane for the negative triangle $G$.
Let us call vertex $p_0$ and vertex $p_a$ the two vertices that are connected by a negative edge but are not part of the triangle of negative edges. 
Without loss of generality, let us assume that $D(p_0) = (0,0)$ and $D(p_a) = (0,a)$. 
Since the edge that connects $p_0$ with $p_a$ is negative, every vertex that is connected with $p_0$ via a positive edge has to be placed
in the interior of the circumference centered at $(0,0)$ and with radius $a$ (the distance between $p_0$ and $p_a$). 


Equivalently, every vertex connected to $p_a$ via a positive edge has to be placed 
in the interior of the circumference centered at point $(0,a)$ and with radius $a$. Therefore, since the remaining three vertices in $G$ are connected by positive edges to $p_0$ and $p_a$, the three of them 
have to be placed in the intersection of the two circumferences described above. The edge that connects $p_0$ with $p_a$ cuts that intersection in two halves. Since there are three other vertices in $G$, at least 
two out of those three are placed in the same half. Without loss of generality, we assume that those two vertices are placed in the half where the $x$ coordinate is positive, and we call them vertices 
$p_1$ and $p_2$, placed at $(x_1,y_1)$ and $(x_2,y_2)$ respectively.
Figure \ref{fig:notationproof2} depicts graph $G$ placed in the plane as described here, and shows the defined notation.

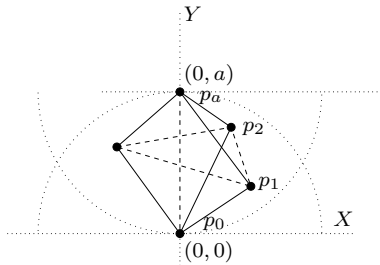
\begin{figure}[h!]
\begin{center}
\begin{picture}(0,0)%
\includegraphics{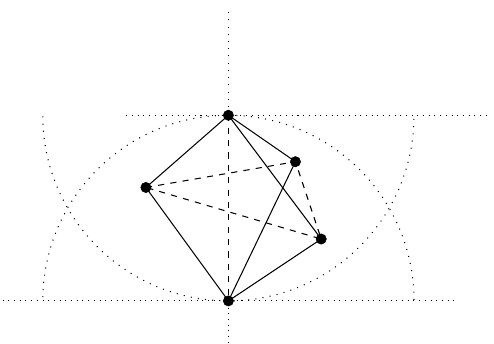}%
\end{picture}%
\setlength{\unitlength}{1302sp}%
\begingroup\makeatletter\ifx\SetFigFont\undefined%
\gdef\SetFigFont#1#2#3#4#5{%
  \reset@font\fontsize{#1}{#2pt}%
  \fontfamily{#3}\fontseries{#4}\fontshape{#5}%
  \selectfont}%
\fi\endgroup%
\begin{picture}(7169,4951)(3279,-8483)
\put(9526,-7711){\makebox(0,0)[lb]{\smash{{\SetFigFont{8}{9.6}{\rmdefault}{\mddefault}{\updefault}{\color[rgb]{0,0,0}$X$}%
}}}}
\put(6676,-3811){\makebox(0,0)[lb]{\smash{{\SetFigFont{8}{9.6}{\rmdefault}{\mddefault}{\updefault}{\color[rgb]{0,0,0}$Y$}%
}}}}
\put(6676,-4936){\makebox(0,0)[lb]{\smash{{\SetFigFont{8}{9.6}{\rmdefault}{\mddefault}{\updefault}{\color[rgb]{0,0,0}$(0,a)$}%
}}}}
\put(6676,-8311){\makebox(0,0)[lb]{\smash{{\SetFigFont{8}{9.6}{\rmdefault}{\mddefault}{\updefault}{\color[rgb]{0,0,0}$(0,0)$}%
}}}}
\put(7801,-5911){\makebox(0,0)[lb]{\smash{{\SetFigFont{8}{9.6}{\rmdefault}{\mddefault}{\updefault}{\color[rgb]{0,0,0}$p_2$}%
}}}}
\put(8101,-6961){\makebox(0,0)[lb]{\smash{{\SetFigFont{8}{9.6}{\rmdefault}{\mddefault}{\updefault}{\color[rgb]{0,0,0}$p_1$}%
}}}}
\put(7051,-7711){\makebox(0,0)[lb]{\smash{{\SetFigFont{8}{9.6}{\rmdefault}{\mddefault}{\updefault}{\color[rgb]{0,0,0}$p_0$}%
}}}}
\put(6976,-5311){\makebox(0,0)[lb]{\smash{{\SetFigFont{8}{9.6}{\rmdefault}{\mddefault}{\updefault}{\color[rgb]{0,0,0}$p_a$}%
}}}}
\end{picture}%
\caption{This figure shows the notation used in the proof. Dashed lines in the graph represent negative edges and continuous lines represent positive edges. 
The signed graph is placed in the Euclidean plane such that $p_0$ is placed at point $(0,0)$ and $p_a$ is placed at point $(0,a)$.}
\label{fig:notationproof2}
\end{center}
\end{figure}

Without loss of generality, we assume $x_1 \geq x_2$. Also, we assume $y_1 \leq a/2$, otherwise we turn $D(G)$
along the axis $Y=a/2$. Then, the square of the distance between $p_1$ and $p_2$ is equal to $(x_1-x_2)^2 + (y_1-y_2)^2$, while the square of the distance 
between $p_1$ and $p_a$ is equal to $x_1^2 + (y_1-a)^2$. Let us take the difference
between those distances and prove that $p_1$ is closer to $p_2$ than to $p_a$. 
\begin{eqnarray*}
d(D(p_1),D(p_a))^2-d(D(p_1),D(p_2))^2&=& x_1^2 + (y_1-a)^2 - (x_1-x_2)^2 - (y_1-y_2)^2 \\
&=& 2x_1x_2 - x_2^2 - y_2^2 +2y_1y_2 + a^2 - 2ay_1 \\
&=&2x_1x_2 - x_2^2 + (a-y_2)(y_2+a-2y_1)\\
&\geq&0
\end{eqnarray*}
The inequality comes from: $x_1\geq x_2$, then $2x_1x_2 - x_2^2\geq 0$; and from  $a\geq y_2$ and $a/2\geq y_1$, then  $(a-y_2)(y_2+a-2y_1)\geq0$.
Therefore, $p_1$ is closer to $p_2$ than to $p_a$, which is a contradiction because the edge that connects $p_1$ and $p_2$ is negative and the edge 
that connects $p_1$ and $p_a$ is positive, both edges being adjacent to $p_1$. Hence, $D(G)$ does not exist. 
\end{proof}
\paragraph{Negative cluster} The \emph{negative cluster} is a signed graph with seven vertices. Six vertices out of the seven
are connected in a cluster of negative edges. The seventh vertex, called central vertex, is connected via a positive edge 
to each of the six vertices in the cluster. Figure \ref{counterexampledim2}\textit{(b)} depicts the negative cluster. 
\begin{lemma}\label{lem:negativecluster}
The negative cluster is a signed graph without valid drawing in $\mathbb{R}^2$.
\end{lemma}
\begin{proof} 
The proof is by contradiction. Let us assume that there exists a valid drawing $D(G)$ for the negative cluster in the Euclidean plane. 
Let us denote the central vertex by $p_0$ and let us depict in $D(G)$ a small enough circumference centered at $p_0$ such that it does not contain any other vertex in it. 
Since the circumference does not contain any other vertex in it, every edge that connects $p_0$ with any other vertex looks like a radius of the circumference. Due to the fact that 
there are six radii, there are at least two clockwise consecutive of them that create an angle of size at most $\pi/3$. 
We call the two vertices at the end of these two edges $p_a$ and $p_b$.  
Without loss of generality, we assume that  $d(D(p_0),D(p_a)) \leq d(D(p_0),D(p_b))$.
We prove, then, that $d(D(p_a),D(p_b)) \leq d(D(p_0),D(p_b))$. Concluding the contradiction 
since the edge $(p_0,p_b)$ is positive and the edge $(p_a,p_b)$ is negative.


Since $D(G)$ is a valid drawing, it holds: $d(D(p_0),D(p_a)) < d(D(p_a),D(p_b))$ and $d(D(p_0),D(p_b))<d(D(p_a),D(p_b))$. 
The proof now proceeds using trigonometric tools in the triangle $p_0p_ap_b$. We denote $\alpha$ the angle at vertex $p_0$. 
We remind you that $\alpha \leq \pi/3$. We denote $\beta$ the angle at vertex $p_b$. Let us denote $l$ the altitude of the 
triangle that is perpendicular to edge $(p_0,p_b)$. Let us denote $b'$ the intersection point between $l$ and the edge $(p_0,p_b)$.
Figure \ref{fig:notation} depicts the described notation. 

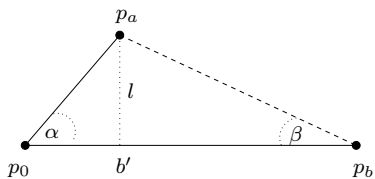
\begin{figure}[h!]
\begin{center}
\begin{picture}(0,0)%
\includegraphics{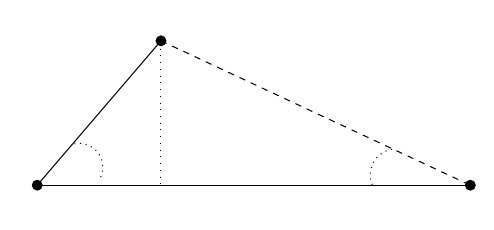}%
\end{picture}%
\setlength{\unitlength}{1302sp}%
\begingroup\makeatletter\ifx\SetFigFont\undefined%
\gdef\SetFigFont#1#2#3#4#5{%
  \reset@font\fontsize{#1}{#2pt}%
  \fontfamily{#3}\fontseries{#4}\fontshape{#5}%
  \selectfont}%
\fi\endgroup%
\begin{picture}(7155,3318)(3361,-7600)
\put(5851,-6061){\makebox(0,0)[lb]{\smash{{\SetFigFont{8}{9.6}{\rmdefault}{\mddefault}{\updefault}{\color[rgb]{0,0,0}$l$}%
}}}}
\put(5626,-4561){\makebox(0,0)[lb]{\smash{{\SetFigFont{8}{9.6}{\rmdefault}{\mddefault}{\updefault}{\color[rgb]{0,0,0}$p_a$}%
}}}}
\put(10150,-7486){\makebox(0,0)[lb]{\smash{{\SetFigFont{8}{9.6}{\rmdefault}{\mddefault}{\updefault}{\color[rgb]{0,0,0}$p_b$}%
}}}}
\put(3576,-7486){\makebox(0,0)[lb]{\smash{{\SetFigFont{8}{9.6}{\rmdefault}{\mddefault}{\updefault}{\color[rgb]{0,0,0}$p_0$}%
}}}}
\put(8926,-6886){\makebox(0,0)[lb]{\smash{{\SetFigFont{8}{9.6}{\rmdefault}{\mddefault}{\updefault}{\color[rgb]{0,0,0}$\beta$}%
}}}}
\put(4276,-6811){\makebox(0,0)[lb]{\smash{{\SetFigFont{8}{9.6}{\rmdefault}{\mddefault}{\updefault}{\color[rgb]{0,0,0}$\alpha$}%
}}}}
\put(5626,-7486){\makebox(0,0)[lb]{\smash{{\SetFigFont{8}{9.6}{\rmdefault}{\mddefault}{\updefault}{\color[rgb]{0,0,0}$b'$}%
}}}}
\end{picture}%
\caption{This figure shows the notation used in the proof. Dashed line in triangle $p_0p_ap_b$ represents a negative edge and continuous lines represent positive edges. 
Line $l$ represents the altitude of triangle $p_0p_ap_b$.} 
\label{fig:notation}
\end{center}
\end{figure}

%

Using basic trigonometry, we have: 

\begin{equation}\label{eq:d(p_0,p_a)}
d(D(p_0),D(p_a)) = \frac{\sin \beta}{\sin \alpha}d(D(p_a),D(p_b))
\end{equation}
On the other hand, by definition of cosine, it holds: $$d(D(p_0),D(p_a))\cos \alpha = d(D(p_0),b')$$ and $$d(D(p_a),D(p_b)) \cos \beta = d(D(p_b),b').$$ 

Now, applying the fact that $d(D(p_0),b') + d(D(p_b),b') = d(D(p_0),D(p_b))$ in the inequality $d(D(p_0),D(p_b))<d(D(p_a),D(p_b))$, we obtain: 
\begin{eqnarray*}
d(D(p_a),D(p_b)) &>& d(D(p_0),b') + d(D(p_b),b') \\
d(D(p_a),D(p_b)) &>& d(D(p_0),D(p_a))\cos \alpha + d(D(p_a),D(p_b)) \cos \beta \\
d(D(p_a),D(p_b))&>&\frac{\sin \beta}{\sin \alpha}d(D(p_a),D(p_b))\cos \alpha +  d(D(p_a),D(p_b)) \cos \beta, 
\end{eqnarray*}
where the last inequality comes from Equation \ref{eq:d(p_0,p_a)}. Therefore, it holds: $$1 > \frac{\sin \beta}{\sin \alpha}\cos \alpha +  \cos \beta.$$
Since $\alpha \leq \pi/3$ and the cotangent function is monotone decreasing in the interval $[0,\pi/3]$, it holds that $\frac{\cos \alpha}{\sin \alpha} \geq \frac{1}{\sqrt{3}}$.
Hence, it also holds $1 > \frac{\sin \beta}{\sqrt{3}} +  \cos \beta$. The last inequality implies that $\beta > \pi/3$. 

On the other hand, using trigonometry we obtain as well: 
 \begin{equation}\label{eq:altitude}
d(D(p_0),D(p_a))\sin\alpha=d(D(p_a),D(p_b))\sin\beta=|l|.
\end{equation}
Since $d(D(p_0),D(p_a))<d(D(p_a),D(p_b))$, from Equation (\ref{eq:altitude}) we obtain $\sin\alpha>\sin\beta$.
Since sine is a monotone increasing function in the interval $[0,\pi/3]$, it holds $\alpha > \beta$.

Therefore, it can not occur at the same time $d(D(p_0),D(p_a))<d(D(p_a),D(p_b))$ and $d(D(p_0),D(p_b))<d(D(p_a),D(p_b))$. Since we asume that 
$d(D(p_0),D(p_a))\leq d(D(p_0),D(p_b))$, we can conclude that  $d(D(p_a),D(p_b)) \leq d(D(p_0),D(p_b))$, which is a contradiction. 
\end{proof}

\subsection{Signed graphs without valid drawing in the line}
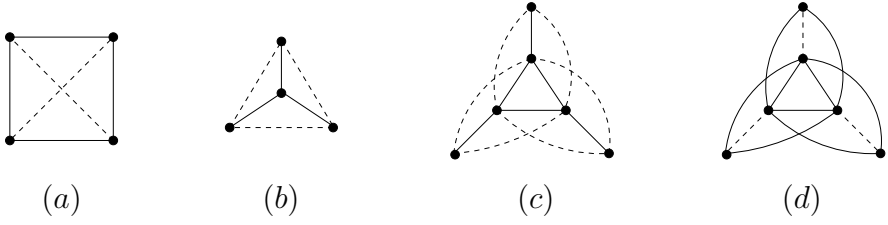
\begin{figure}[t!]
\begin{center}
\begin{picture}(0,0)(0,0)%
\includegraphics[width= 0.9\textwidth]{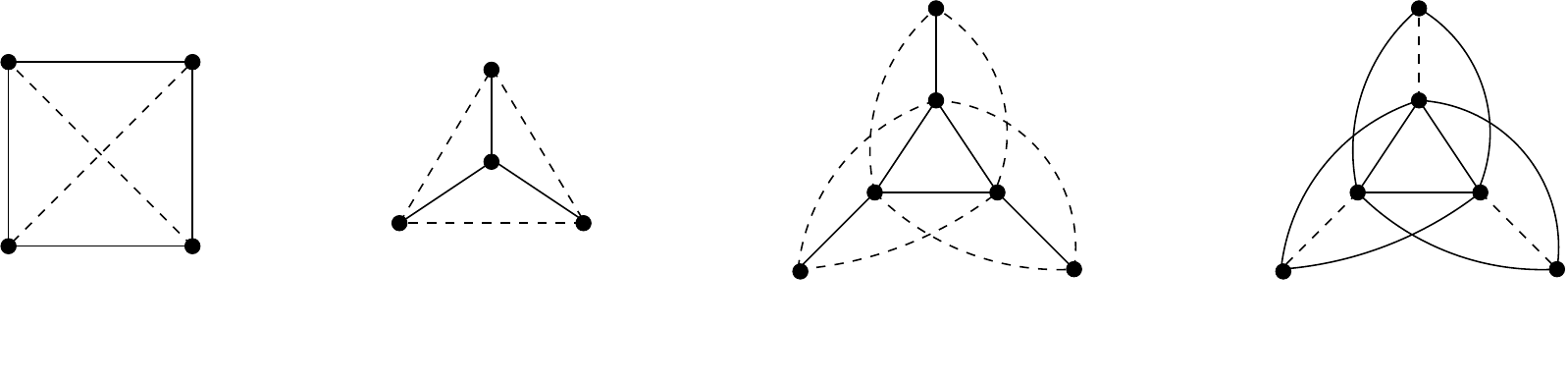}%
\end{picture}%
\setlength{\unitlength}{1973sp}%
\begingroup\makeatletter\ifx\SetFigFont\undefined%
\gdef\SetFigFont#1#2#3#4#5{%
  \reset@font\fontsize{#1}{#2pt}%
  \fontfamily{#3}\fontseries{#4}\fontshape{#5}%
  \selectfont}%
\fi\endgroup%
\begin{picture}(15316,3716)(-82,-6469)
\put(9676,-6361){\makebox(0,0)[lb]{\smash{{\SetFigFont{12}{14.4}{\familydefault}{\mddefault}{\updefault}{\color[rgb]{0,0,0}$(d)$}%
}}}}
\put(376,-6361){\makebox(0,0)[lb]{\smash{{\SetFigFont{12}{14.4}{\familydefault}{\mddefault}{\updefault}{\color[rgb]{0,0,0}$(a)$}%
}}}}
\put(3151,-6361){\makebox(0,0)[lb]{\smash{{\SetFigFont{12}{14.4}{\familydefault}{\mddefault}{\updefault}{\color[rgb]{0,0,0}$(b)$}%
}}}}
\put(6351,-6361){\makebox(0,0)[lb]{\smash{{\SetFigFont{12}{14.4}{\familydefault}{\mddefault}{\updefault}{\color[rgb]{0,0,0}$(c)$}%
}}}}
\end{picture}%
\caption{This figure shows four signed graphs, where each of these four signed graphs belongs to one of the four families of signed graphs without valid drawing in the Euclidean line. Dashed lines represent negative edges and continuous lines represent positive edges. 
Subfigure \textit{(a)} depicts signed graph $\mathcal{F}_1(4,2)$. Subfigure \textit{(b)} depicts signed graph $\mathcal{F}_2(3)$. 
Subfigure \textit{(c)} depicts signed graph $\mathcal{F}_3(3)$. Subfigure \textit{(d)} depicts signed graph $\mathcal{F}_4(3)$.}
\label{counterexampledim1}
\end{center}
\end{figure}

We now proceed to present the four infinite families of signed graphs without valid drawing in the Euclidean line. We first define each of the families and then prove our results.  
\paragraph{Description of the families} For simplicity, we denote the families with letters $\mathcal{F}_1$, $\mathcal{F}_2$, $\mathcal{F}_3$ and $\mathcal{F}_4$. 
\begin{itemize}
\item[*] \emph{Family $\mathcal{F}_1$:} A signed graph in the family $\mathcal{F}_1$ consists of $n$ vertices connected in a cycle of positive edges; plus negative edges that connect each pair of vertices separated at distance exactly $k$ in the cycle. 
Values $n$ and $k$ determine any signed graph that belongs to the family. Hence, we use $\mathcal{F}_1(n,k)$ to denote the signed graph in family $\mathcal{F}_1$ and consists of $n$ vertices and  
where each negative edge connects a pair of vertices separated at distance exactly $2 \leq k \leq n/2$ in the cycle. A graphic representation of $\mathcal{F}_1(4,2)$ is shown in Figure \ref{counterexampledim1}\textit{(a)}.

\item[*] \emph{Family $\mathcal{F}_2$:} A signed graph in the family $\mathcal{F}_2$ consists of $n$ vertices connected by a cycle of negative edges; plus a central vertex connected via a positive edge to each of the $n$ vertices in the cycle. 
A signed graph that belongs to the family is fully determined by the size of the negative cycle. Hence, we denote by $\mathcal{F}_2(n)$ the signed graph that belongs to the family $\mathcal{F}_2$ and consists of $n$ vertices plus one central vertex. A graphic representation of $\mathcal{F}_2(3)$ is shown in Figure ~\ref{counterexampledim1}\textit{(b)}.

\item[*] \emph{Family $\mathcal{F}_3$:} A signed graph in the family $\mathcal{F}_3$ consists of $2n$ vertices, $2n$ positive edges and $2n$ negative edges. A cycle of positive edges connects $n$ out of the $2n$ vertices. 
Each of the remaining vertices is matched with one of the vertices in the cycle via a positive edge and connected via a negative edge to the preceding and the following vertices of the matched vertex in the cycle.   
The number of vertices in the cycle fully determines a signed graph that belongs to the family. Hence, we denote  $\mathcal{F}_3(n)$ the signed graph that belongs to the family $\mathcal{F}_3$
 and consists of $2n$ vertices. A graphic representation of $\mathcal{F}_3(3)$ is shown in Figure~\ref{counterexampledim1}\textit{(c)}.

\item[*] \emph{Family $\mathcal{F}_4$:} A signed graph in the family $\mathcal{F}_4$ consists of $2n$ vertices, $3n$ positive edges and $n$ negative edges. A cycle of positive edges connects $n$ out of the $2n$ vertices. 
Each of the remaining vertices is matched with one of the vertices in the cycle via a negative edge and connected via a positive edge to the preceding and the following vertices of the matched vertex in the cycle.   
The number of vertices in the cycle fully determines a signed graph that belongs to the family. Hence, we denote  $\mathcal{F}_4(n)$ the signed graph that belongs to the family $\mathcal{F}_4$ and
 consists of $2n$ vertices. A graphic representation of $\mathcal{F}_4(3)$ is shown in Figure~\ref{counterexampledim1}\textit{(d)}.

\end{itemize}

%
\begin{lemma}\label{lem:positivesquare}
For any $n$ and $k$ such that $2\leq k \leq n/2$, $\mathcal{F}_1(n,k)$ is a signed graph without valid drawing in the Euclidean line.
\end{lemma}
\begin{proof} 
The proof of this Lemma relies in the definition of $k$-central vertex and Lemma 2 presented in \cite{minsa}.  A $k$-central vertex can be seen 
as a vertex that together with other $2k$ vertices induce a subgraph equivalent to the one induced by any vertex in the cycle together with its $k$ first neighbors at its left and right hand side in the cycle. 
On the other hand, Lemma 2 presented in \cite{minsa} shows that a central vertex can not be placed as the leftmost or rightmost vertex in a valid drawing. Since any valid drawing 
needs a leftmost and a rightmost vertex, then it can not exist a valid drawing for $\mathcal{F}_1(n,k)$. 
\end{proof} 
%
\begin{lemma}\label{lem:positivestar}
For any $n\geq 1$, $\mathcal{F}_2(2n-1)$ is a signed graph without valid drawing in the Euclidean line.
\end{lemma}
\begin{proof}
The proof is  by contradiction. Let us assume that $\mathcal{F}_2(2n-1)$ has a valid drawing in the line, and call it $D(\mathcal{F}_2(2n-1))$. 
Since the cycle has an odd number of vertices, by The Pigeon Hole Principle, in any drawing of $\mathcal{F}_2(2n-1)$ in the line, 
two consecutive vertices of the negative cycle are placed at the same side of the central vertex, either its left or its right hand side.
In particular, that is true for $D(\mathcal{F}_2(2n-1))$. 
Without loss of generality, we assume that in $D(\mathcal{F}_2(2n-1))$ there are two consecutive vertices of the negative cycle 
placed at the right hand side of the central vertex. Let us call $p_0$ the central vertex, $p_1$ the first vertex of the two consecutive vertices at 
the right hand side of $p_0$, and $p_2$ the second of the two consecutive vertices, from left to right. 
Hence, the contradiction comes from the fact that vertices $p_0$ and $p_2$ are connected by a positive edge, 
while vertices $p_1$ and $p_2$ are connected by a negative edge. Therefore, $p_2$ generates the contradiction because 
the positive edge that connects vertices $p_0$ and $p_2$ is longer that the negative edge that connects $p_1$ and $p_2$. 
\end{proof}

The proof of the following three lemmas is based in a common remark. 
\emph{Consider any embedding of   
a cycle of odd length in the line. There exists at least one vertex such that one neighbor is embedded at its left side and the other neighbor is embedded at its right hand side. 
}
A simple argument that proves this affirmation follows. Assume that there exists an embedding of a cycle of odd length in the line such that for every vertex, 
it holds that its two neighbors are embedded at the same side, either left or right hand side. Then, pick any vertex and start a walk in the cycle. 
The walk according to the embedding will cross edges consecutively going from left to right and then from right to left one by one. Let us label with an L the edges 
crossed from left to right and with an R the edges crossed from right to left. The labeling for the edges of the cycle obtained according to the walk that follows
the embedding, produces iteratively L's and R's. Hence, there is the same number of L's and R's, which is a contradiction since the length of the cycle is odd. 

\begin{lemma}\label{lem:positivetriangle}  
For any $n\geq 1$, $\mathcal{F}_3(2n-1)$ is a signed graph without valid drawing in the Euclidean line.
\end{lemma}
\begin{proof}
The proof is by contradiction. Let us start from a valid drawing $D(\mathcal{F}_3(2n-1))$. According the previous remark 
there exists one special vertex such that $D(\mathcal{F}_3(2n-1))$ embeds one of its neighbors at its left side and the other neighbor at its right hand side. 
%
Let us denote that special vertex $p_c$. Let us denote as well by $p_l$ and $p_r$ its neighbors embedded at its left and right hand side, respectively. 
Let us denote by $\hat{p}_c$ the vertex matched with $p_c$. Vertex $\hat{p}_c$ is connected via a negative edge to $p_l$ and to $p_r$. 
Let us assume that $D(\mathcal{F}_3(2n-1))$ places $\hat{p}_c$ at the left of $p_c$. Then, vertex $\hat{p}_c$ is either between $p_l$ and $p_c$ or it is 
embedded at the left of $p_l$. If vertex $\hat{p}_c$ is embedded between $p_l$ and $p_c$, there is a contradiction since the negative edge $(p_l,\hat{p}_c)$ would be shorter than 
the positive edge $(p_l,p_c)$.  On the other hand, if vertex $\hat{p}_c$ is 
embedded at the left side of $p_l$,  there is a contradiction since the negative edge $(\hat{p}_c, p_l)$ would be shorter than 
the positive edge $(\hat{p}_c,p_c)$. 
The argument is equivalent when vertex $\hat{p}_c$ is embedded at the right hand side of $p_c$, but in that case it is constructed using $p_r$. 
\end{proof}

\begin{lemma}\label{lem:d-family}  
For any $n\geq 1$, $\mathcal{F}_4(2n-1)$ is a signed graph without valid drawing in the Euclidean line.
\end{lemma}
\begin{proof}
The proof is by contradiction. Let us start from a valid drawing $D(\mathcal{F}_4(2n-1))$. According the previous remark 
there exists one special vertex such that $D(\mathcal{F}_4(2n-1))$ embeds one of its neighbors at its left side and the other neighbor at its right hand side. 
Let us use the notation as in the previous proof for vertices $p_c$, $p_l$, $p_r$ and $\hat{p}_c$. 
 Vertex $\hat{p}_c$ is connected via a positive edge to $p_l$ and to $p_r$ and via a negative edge to $p_c$. 
Let us assume that $D(\mathcal{F}_4(2n-1))$ embeds $\hat{p}_c$ at the left of $p_c$. 
Then, the negative edge $(\hat{p}_c, p_c)$ is shorter than the positive edge $(\hat{p}_c, p_r)$. Equivalently, 
if we assume that $D(\mathcal{F}_4(2n-1))$ embeds $\hat{p}_c$ at the right hand side of $p_c$, the argument is constructed using $p_l$ instead of $p_r$. 
\end{proof}

\subsection{Minimal graphs without valid drawing} 
An interesting remark about graphs with valid drawing in a given dimension 
is the fact that the property of having a valid drawing 
is heritable through induced subgraphs. 
Equivalently, previous remark can be stated as follows. 
\emph{Let $G$ be a signed graph. $G$ is a signed graph without valid drawing in $\mathbb{R}^\ell$ if and only if 
there exists an induced subgraph of $G$ without valid drawing in $\mathbb{R}^\ell$. }
Pushing the previous remark to the extreme case when the only induced subgraph without valid drawing is the original signed graph itself, 
we obtain the definition of \emph{minimal} graph without valid drawing. 
\begin{definition}\label{def:minimal}
Let $G$ be a signed graph without valid drawing in $\mathbb{R}^\ell$. The signed graph $G$ is called \emph{minimal} without valid drawing
if and only if every proper induced subgraph of $G$ has a valid drawing in 
$\mathbb{R}^\ell$.  
\end{definition}

\begin{figure}[t!]
\begin{center}
\begin{picture}(0,0)%
\includegraphics{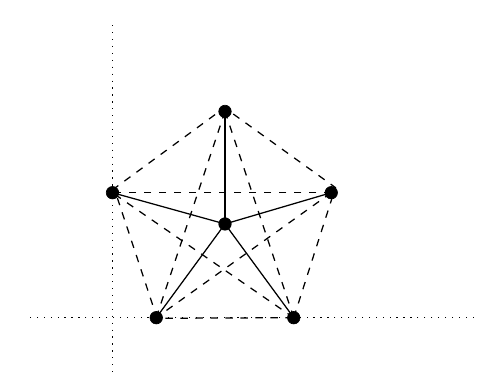}%
\end{picture}%
\setlength{\unitlength}{1579sp}%
\begingroup\makeatletter\ifx\SetFigFont\undefined%
\gdef\SetFigFont#1#2#3#4#5{%
  \reset@font\fontsize{#1}{#2pt}%
  \fontfamily{#3}\fontseries{#4}\fontshape{#5}%
  \selectfont}%
\fi\endgroup%
\begin{picture}(5722,4510)(2101,-4583)
\put(4876,-1186){\makebox(0,0)[lb]{\smash{{\SetFigFont{10}{12.0}{\rmdefault}{\mddefault}{\updefault}{\color[rgb]{0,0,0}$p_4$}%
}}}}
\put(5776,-4261){\makebox(0,0)[lb]{\smash{{\SetFigFont{10}{12.0}{\rmdefault}{\mddefault}{\updefault}{\color[rgb]{0,0,0}$p_1$}%
}}}}
\put(3976,-4336){\makebox(0,0)[lb]{\smash{{\SetFigFont{10}{12.0}{\rmdefault}{\mddefault}{\updefault}{\color[rgb]{0,0,0}$p_2$}%
}}}}
\put(2701,-2536){\makebox(0,0)[lb]{\smash{{\SetFigFont{10}{12.0}{\rmdefault}{\mddefault}{\updefault}{\color[rgb]{0,0,0}$p_3$}%
}}}}
\put(4876,-2986){\makebox(0,0)[lb]{\smash{{\SetFigFont{10}{12.0}{\rmdefault}{\mddefault}{\updefault}{\color[rgb]{0,0,0}$p_6$}%
}}}}
\put(2101,-4336){\makebox(0,0)[lb]{\smash{{\SetFigFont{10}{12.0}{\rmdefault}{\mddefault}{\updefault}{\color[rgb]{0,0,0}$(0,0)$}%
}}}}
\put(3601,-361){\makebox(0,0)[lb]{\smash{{\SetFigFont{10}{12.0}{\rmdefault}{\mddefault}{\updefault}{\color[rgb]{0,0,0}$F_2'$}%
}}}}
\put(6676,-2536){\makebox(0,0)[lb]{\smash{{\SetFigFont{10}{12.0}{\rmdefault}{\mddefault}{\updefault}{\color[rgb]{0,0,0}$p_5$}%
}}}}
\end{picture}%
\caption{This figure depicts valid drawings in the Euclidean plane for subgraphs of the negative cluster signed graph. 
Dashed lines represent negative edges, while continuous lines represent positive edges.}
\label{fig:validedrawingnegativecluster}
\end{center}
\end{figure}
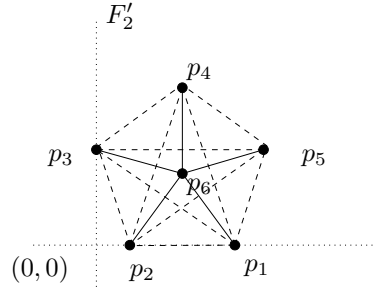

%
Let us denote $\mathcal{M}^\ell$ the set of all minimal graphs without a valid drawing in $\mathbb{R}^\ell$. 
We consider the problem 
of characterize the set $\mathcal{M}^\ell$.
The interest of this problem arises in the search of patterns 
that create problems when searching for a valid drawing of a signed graph. 
Note that, a signed graph $G$ has not a valid drawing in $\mathbb{R}^\ell$ if and only if $G$ does contain 
as induced subgraph a $G'$ in $\mathcal{M}^\ell$. Our contribution to this problem is the fact 
that every signed graph described previously in this section is minimal.  Hence they belong to either  
$\mathcal{M}^1$ or $\mathcal{M}^2$, respectively.
\begin{theorem}
Let $F_1$ be the negative triangle signed graph and $F_2$ be the cluster signed graph, then $\{F_1,F_2\}$ is a subset of $\mathcal{M}^2$.
\end{theorem}
\begin{proof}
In order to prove the Theorem, we show valid drawings for induced subgraphs of $F_1$ and $F_2$. 
We start with the negative triangle $F_1$. By symmetry, $F_1$ has two different induced subgraphs obtained by deleting only one vertex. The first subgraph is 
obtained by deleting one vertex of the triangle composed by negative edges, we call it $F_1'$. The second subgraph is obtained by deleting vertex 
that is not in the triangle of negative edges, we call it $F_1''$. 
%
%
\begin{figure}[t!]
\begin{center}
\begin{picture}(0,0)%
\includegraphics{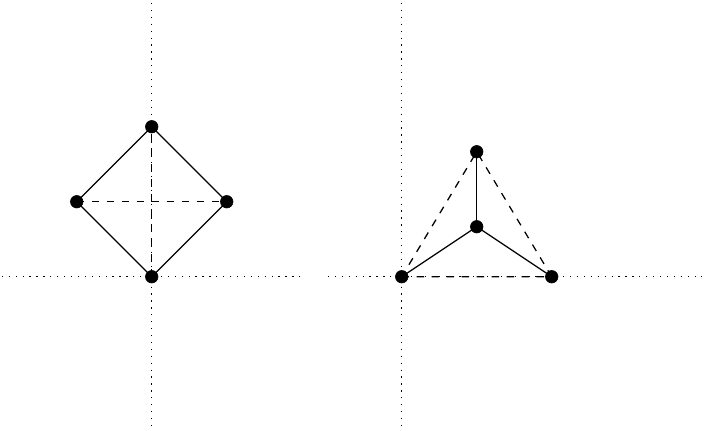}%
\end{picture}%
\setlength{\unitlength}{1579sp}%
\begingroup\makeatletter\ifx\SetFigFont\undefined%
\gdef\SetFigFont#1#2#3#4#5{%
  \reset@font\fontsize{#1}{#2pt}%
  \fontfamily{#3}\fontseries{#4}\fontshape{#5}%
  \selectfont}%
\fi\endgroup%
\begin{picture}(8444,5144)(2379,-4883)
\put(9001,-3586){\makebox(0,0)[lb]{\smash{{\SetFigFont{10}{12.0}{\rmdefault}{\mddefault}{\updefault}{\color[rgb]{0,0,0}$p_3$}%
}}}}
\put(4726,-2686){\makebox(0,0)[lb]{\smash{{\SetFigFont{10}{12.0}{\rmdefault}{\mddefault}{\updefault}{\color[rgb]{0,0,0}$p_3$}%
}}}}
\put(2626,-2236){\makebox(0,0)[lb]{\smash{{\SetFigFont{10}{12.0}{\rmdefault}{\mddefault}{\updefault}{\color[rgb]{0,0,0}$p_4$}%
}}}}
\put(3151,-3436){\makebox(0,0)[lb]{\smash{{\SetFigFont{10}{12.0}{\rmdefault}{\mddefault}{\updefault}{\color[rgb]{0,0,0}$(0,0)$}%
}}}}
\put(6076,-3436){\makebox(0,0)[lb]{\smash{{\SetFigFont{10}{12.0}{\rmdefault}{\mddefault}{\updefault}{\color[rgb]{0,0,0}$(0,0)$}%
}}}}
\put(7276,-3586){\makebox(0,0)[lb]{\smash{{\SetFigFont{10}{12.0}{\rmdefault}{\mddefault}{\updefault}{\color[rgb]{0,0,0}$p_1$}%
}}}}
\put(4276,-3586){\makebox(0,0)[lb]{\smash{{\SetFigFont{10}{12.0}{\rmdefault}{\mddefault}{\updefault}{\color[rgb]{0,0,0}$p_1$}%
}}}}
\put(7426,-61){\makebox(0,0)[lb]{\smash{{\SetFigFont{10}{12.0}{\rmdefault}{\mddefault}{\updefault}{\color[rgb]{0,0,0}$F_1''$}%
}}}}
\put(2626,-61){\makebox(0,0)[lb]{\smash{{\SetFigFont{10}{12.0}{\rmdefault}{\mddefault}{\updefault}{\color[rgb]{0,0,0}$F_1'$}%
}}}}
\put(7651,-1261){\makebox(0,0)[lb]{\smash{{\SetFigFont{10}{12.0}{\rmdefault}{\mddefault}{\updefault}{\color[rgb]{0,0,0}$p_2$}%
}}}}
\put(4351,-1111){\makebox(0,0)[lb]{\smash{{\SetFigFont{10}{12.0}{\rmdefault}{\mddefault}{\updefault}{\color[rgb]{0,0,0}$p_2$}%
}}}}
\put(8251,-2386){\makebox(0,0)[lb]{\smash{{\SetFigFont{10}{12.0}{\rmdefault}{\mddefault}{\updefault}{\color[rgb]{0,0,0}$p_4$}%
}}}}
\end{picture}
\caption{This figure depicts valid drawings in the Euclidean plane for two induced subgraphs of the negative triangle signed graph. 
Dashed lines represent negative edges, while continuous lines represent positive edges.}
\label{fig:validrawingnegativetrianlge}
\end{center}
\end{figure}
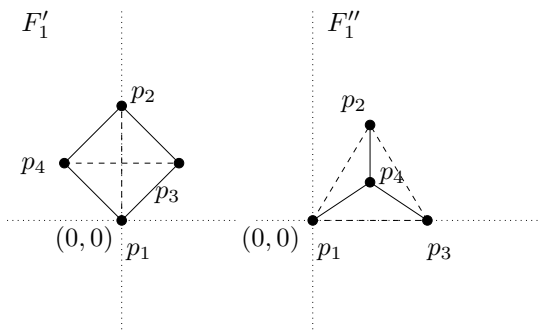

In order to complete the proof for $F_1$, we give positions for every vertex in the two cases $F_1'$ and $F_1''$. 
Two valid drawings for $F_1'$ and $F_1''$ are depicted in Figure \ref{fig:validrawingnegativetrianlge}. These two draiwngs follow: 
\begin{eqnarray*}
D(F_1')&=&\{(0,0);(0,1);(1/2,1/2);(-1/2,1/2)\}, \\
D(F_1'')&=&\{(0,0);(1/2,\sqrt{3}/2);(1,0);(1/2,\sqrt{3}/4)\},\\
\end{eqnarray*}

We proceed equivalently with the negative cluster signed graph $F_2$. By symmetry, $F_2$ has two different induced subgraphs obtained by deleting one vertex. 
First $F_2'$ is obtained by deleting a vertex of the negative cluster, while $F_2''$ is obtained by deleting the central vertex. 
Note that, if the central vertex is deleted, as result we obtain a signed graph with negative edges only. Hence, any drawing is valid. 
In order to conclude the proof, we give valid drawing for $F_2'$. 
\begin{eqnarray*}
D(F_2')=&&\{\left(\frac{1}{4}(\sqrt{5}+3),0\right);\left(\frac{1}{4}(\sqrt{5}-1),0\right);\left(0,\sqrt{\frac{5}{8}+\frac{\sqrt{5}}{8}}\right);
\\&&\left(\frac{1}{4}(\sqrt{5}+1),\frac{1}{2}(1+\sqrt{5})\sqrt{\frac{2}{5-\sqrt{5}}}\right);\left(\frac{1}{2}(\sqrt{5}+1),\sqrt{\frac{5}{8}+\frac{\sqrt{5}}{8}}\right);
\\&&\left(\frac{1}{4}(\sqrt{5}+1),\frac{1}{4}(1+\sqrt{5})\sqrt{\frac{2}{5-\sqrt{5}}}\right)\}
\end{eqnarray*}
Fig. \ref{fig:validedrawingnegativecluster} shows a representation of the valid drawings given here. 
\end{proof} 
%
%
\begin{theorem}
Signed graphs $\mathcal{F}_1(n,k)$, $\mathcal{F}_2(2n-1)$, $\mathcal{F}_3(2n-1)$ and $\mathcal{F}_4(2n-1)$ 
belong to $\mathcal{M}^1$ for all $n \geq 1$ and for all $2\leq k \leq n/2$.
\end{theorem}
\begin{proof}
The proof of this Theorem consists in present valid drawings for any proper induced subgraph of  
$\mathcal{F}_1(n,k)$, $\mathcal{F}_2(2n-1)$, $\mathcal{F}_3(2n-1)$ and $\mathcal{F}_4(2n-1)$ 
for all $n \geq 1$ and for all $2\leq k \leq n/2$. 

Let us start with graph $\mathcal{F}_1(n,k)$, for all $n \geq 1$ and for all $2\leq k \leq n/2$.
We first point out the fact that by symmetry, when one vertex of $\mathcal{F}_1(n,k)$ is deleted, the induced subgraph that we obtain is always the same, independent of the deleted vertex.  
We name from $1$ to $n$ the vertices of $\mathcal{F}_1(n,k)$ clockwise according to the positive cycle, and (w.l.o.g.) we delete vertex $n$. A valid drawing for the resultant signed graph is the following: $D(i)=i$ for $1\leq i < n$. See Figure \ref{fig:validdrawinginducedsubgraphline} for an example of such a drawing applied to the induced subgraph of $\mathcal{F}_1(9,4)$. 

 \begin{figure}[t!]
\begin{center}
 \begin{picture}(0,0)%
\includegraphics{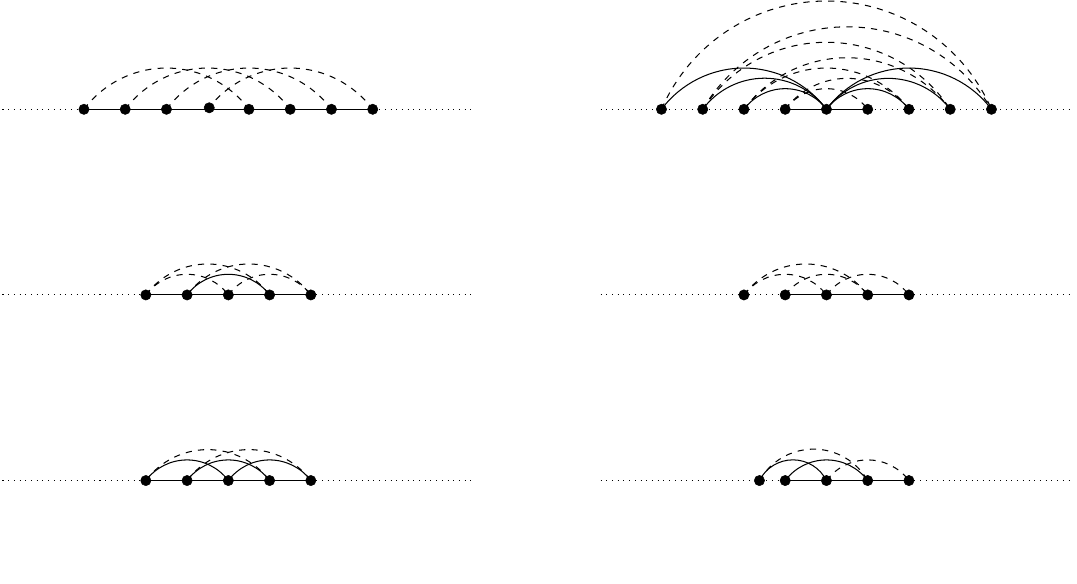}%
\end{picture}%
\setlength{\unitlength}{1302sp}%
\begingroup\makeatletter\ifx\SetFigFont\undefined%
\gdef\SetFigFont#1#2#3#4#5{%
  \reset@font\fontsize{#1}{#2pt}%
  \fontfamily{#3}\fontseries{#4}\fontshape{#5}%
  \selectfont}%
\fi\endgroup%
\begin{picture}(15644,8454)(2379,-9623)
\put(3601,-9511){\makebox(0,0)[lb]{\smash{{\SetFigFont{8}{9.6}{\rmdefault}{\mddefault}{\updefault}{\color[rgb]{0,0,0}$\mathcal{F}_4(3) - \{\hat{p}_1\}$}%
}}}}
\put(5626,-5986){\makebox(0,0)[lb]{\smash{{\SetFigFont{7}{8.4}{\rmdefault}{\mddefault}{\updefault}{\color[rgb]{0,0,0}$p_1$}%
}}}}
\put(6226,-5986){\makebox(0,0)[lb]{\smash{{\SetFigFont{7}{8.4}{\rmdefault}{\mddefault}{\updefault}{\color[rgb]{0,0,0}$p_2$}%
}}}}
\put(5026,-5986){\makebox(0,0)[lb]{\smash{{\SetFigFont{7}{8.4}{\rmdefault}{\mddefault}{\updefault}{\color[rgb]{0,0,0}$p_3$}%
}}}}
\put(6826,-5986){\makebox(0,0)[lb]{\smash{{\SetFigFont{7}{8.4}{\rmdefault}{\mddefault}{\updefault}{\color[rgb]{0,0,0}$\hat{p}_2$}%
}}}}
\put(4426,-5986){\makebox(0,0)[lb]{\smash{{\SetFigFont{7}{8.4}{\rmdefault}{\mddefault}{\updefault}{\color[rgb]{0,0,0}$\hat{p}_3$}%
}}}}
\put(14926,-5986){\makebox(0,0)[lb]{\smash{{\SetFigFont{7}{8.4}{\rmdefault}{\mddefault}{\updefault}{\color[rgb]{0,0,0}$p_3$}%
}}}}
\put(15526,-5986){\makebox(0,0)[lb]{\smash{{\SetFigFont{7}{8.4}{\rmdefault}{\mddefault}{\updefault}{\color[rgb]{0,0,0}$\hat{p}_3$}%
}}}}
\put(14326,-5986){\makebox(0,0)[lb]{\smash{{\SetFigFont{7}{8.4}{\rmdefault}{\mddefault}{\updefault}{\color[rgb]{0,0,0}$p_2$}%
}}}}
\put(13726,-5986){\makebox(0,0)[lb]{\smash{{\SetFigFont{7}{8.4}{\rmdefault}{\mddefault}{\updefault}{\color[rgb]{0,0,0}$\hat{p}_2$}%
}}}}
\put(13126,-5986){\makebox(0,0)[lb]{\smash{{\SetFigFont{7}{8.4}{\rmdefault}{\mddefault}{\updefault}{\color[rgb]{0,0,0}$\hat{p}_1$}%
}}}}
\put(12001,-6811){\makebox(0,0)[lb]{\smash{{\SetFigFont{8}{9.6}{\rmdefault}{\mddefault}{\updefault}{\color[rgb]{0,0,0}$\mathcal{F}_3(3) - \{p_1\}$}%
}}}}
\put(14926,-8686){\makebox(0,0)[lb]{\smash{{\SetFigFont{7}{8.4}{\rmdefault}{\mddefault}{\updefault}{\color[rgb]{0,0,0}$p_3$}%
}}}}
\put(14326,-8686){\makebox(0,0)[lb]{\smash{{\SetFigFont{7}{8.4}{\rmdefault}{\mddefault}{\updefault}{\color[rgb]{0,0,0}$p_2$}%
}}}}
\put(13351,-8686){\makebox(0,0)[lb]{\smash{{\SetFigFont{7}{8.4}{\rmdefault}{\mddefault}{\updefault}{\color[rgb]{0,0,0}$\hat{p}_3$}%
}}}}
\put(15526,-8686){\makebox(0,0)[lb]{\smash{{\SetFigFont{7}{8.4}{\rmdefault}{\mddefault}{\updefault}{\color[rgb]{0,0,0}$\hat{p}_2$}%
}}}}
\put(13726,-8686){\makebox(0,0)[lb]{\smash{{\SetFigFont{7}{8.4}{\rmdefault}{\mddefault}{\updefault}{\color[rgb]{0,0,0}$\hat{p}_1$}%
}}}}
\put(12001,-9511){\makebox(0,0)[lb]{\smash{{\SetFigFont{8}{9.6}{\rmdefault}{\mddefault}{\updefault}{\color[rgb]{0,0,0}$\mathcal{F}_4(3) - \{p_1\}$}%
}}}}
\put(3601,-4261){\makebox(0,0)[lb]{\smash{{\SetFigFont{8}{9.6}{\rmdefault}{\mddefault}{\updefault}{\color[rgb]{0,0,0}$\mathcal{F}_1(9,4) - \{p_9\}$}%
}}}}
\put(12001,-4261){\makebox(0,0)[lb]{\smash{{\SetFigFont{8}{9.6}{\rmdefault}{\mddefault}{\updefault}{\color[rgb]{0,0,0}$\mathcal{F}_2(9) - \{p_9\}$}%
}}}}
\put(3526,-3286){\makebox(0,0)[lb]{\smash{{\SetFigFont{7}{8.4}{\rmdefault}{\mddefault}{\updefault}{\color[rgb]{0,0,0}$p_1$}%
}}}}
\put(4726,-3286){\makebox(0,0)[lb]{\smash{{\SetFigFont{7}{8.4}{\rmdefault}{\mddefault}{\updefault}{\color[rgb]{0,0,0}$p_3$}%
}}}}
\put(5326,-3286){\makebox(0,0)[lb]{\smash{{\SetFigFont{7}{8.4}{\rmdefault}{\mddefault}{\updefault}{\color[rgb]{0,0,0}$p_4$}%
}}}}
\put(5926,-3286){\makebox(0,0)[lb]{\smash{{\SetFigFont{7}{8.4}{\rmdefault}{\mddefault}{\updefault}{\color[rgb]{0,0,0}$p_5$}%
}}}}
\put(6526,-3286){\makebox(0,0)[lb]{\smash{{\SetFigFont{7}{8.4}{\rmdefault}{\mddefault}{\updefault}{\color[rgb]{0,0,0}$p_6$}%
}}}}
\put(7126,-3286){\makebox(0,0)[lb]{\smash{{\SetFigFont{7}{8.4}{\rmdefault}{\mddefault}{\updefault}{\color[rgb]{0,0,0}$p_7$}%
}}}}
\put(7726,-3286){\makebox(0,0)[lb]{\smash{{\SetFigFont{7}{8.4}{\rmdefault}{\mddefault}{\updefault}{\color[rgb]{0,0,0}$p_8$}%
}}}}
\put(4126,-3286){\makebox(0,0)[lb]{\smash{{\SetFigFont{7}{8.4}{\rmdefault}{\mddefault}{\updefault}{\color[rgb]{0,0,0}$p_2$}%
}}}}
\put(11926,-3286){\makebox(0,0)[lb]{\smash{{\SetFigFont{7}{8.4}{\rmdefault}{\mddefault}{\updefault}{\color[rgb]{0,0,0}$p_1$}%
}}}}
\put(12526,-3286){\makebox(0,0)[lb]{\smash{{\SetFigFont{7}{8.4}{\rmdefault}{\mddefault}{\updefault}{\color[rgb]{0,0,0}$p_2$}%
}}}}
\put(13126,-3286){\makebox(0,0)[lb]{\smash{{\SetFigFont{7}{8.4}{\rmdefault}{\mddefault}{\updefault}{\color[rgb]{0,0,0}$p_3$}%
}}}}
\put(13726,-3286){\makebox(0,0)[lb]{\smash{{\SetFigFont{7}{8.4}{\rmdefault}{\mddefault}{\updefault}{\color[rgb]{0,0,0}$p_4$}%
}}}}
\put(14926,-3286){\makebox(0,0)[lb]{\smash{{\SetFigFont{7}{8.4}{\rmdefault}{\mddefault}{\updefault}{\color[rgb]{0,0,0}$p_5$}%
}}}}
\put(15526,-3286){\makebox(0,0)[lb]{\smash{{\SetFigFont{7}{8.4}{\rmdefault}{\mddefault}{\updefault}{\color[rgb]{0,0,0}$p_6$}%
}}}}
\put(16126,-3286){\makebox(0,0)[lb]{\smash{{\SetFigFont{7}{8.4}{\rmdefault}{\mddefault}{\updefault}{\color[rgb]{0,0,0}$p_7$}%
}}}}
\put(16726,-3286){\makebox(0,0)[lb]{\smash{{\SetFigFont{7}{8.4}{\rmdefault}{\mddefault}{\updefault}{\color[rgb]{0,0,0}$p_8$}%
}}}}
\put(14326,-3286){\makebox(0,0)[lb]{\smash{{\SetFigFont{7}{8.4}{\rmdefault}{\mddefault}{\updefault}{\color[rgb]{0,0,0}$p_c$}%
}}}}
\put(3601,-6661){\makebox(0,0)[lb]{\smash{{\SetFigFont{8}{9.6}{\rmdefault}{\mddefault}{\updefault}{\color[rgb]{0,0,0}$\mathcal{F}_3(3) - \{\hat{p}_1\}$}%
}}}}
\put(5626,-8686){\makebox(0,0)[lb]{\smash{{\SetFigFont{7}{8.4}{\rmdefault}{\mddefault}{\updefault}{\color[rgb]{0,0,0}$p_1$}%
}}}}
\put(6226,-8686){\makebox(0,0)[lb]{\smash{{\SetFigFont{7}{8.4}{\rmdefault}{\mddefault}{\updefault}{\color[rgb]{0,0,0}$p_2$}%
}}}}
\put(5026,-8686){\makebox(0,0)[lb]{\smash{{\SetFigFont{7}{8.4}{\rmdefault}{\mddefault}{\updefault}{\color[rgb]{0,0,0}$p_3$}%
}}}}
\put(6826,-8686){\makebox(0,0)[lb]{\smash{{\SetFigFont{7}{8.4}{\rmdefault}{\mddefault}{\updefault}{\color[rgb]{0,0,0}$\hat{p}_3$}%
}}}}
\put(4426,-8686){\makebox(0,0)[lb]{\smash{{\SetFigFont{7}{8.4}{\rmdefault}{\mddefault}{\updefault}{\color[rgb]{0,0,0}$\hat{p}_2$}%
}}}}
\end{picture}%
\caption{This figure depicts valid drawings in the Euclidean line for induced subgraphs of graphs $\mathcal{F}_1(9,4)$, $\mathcal{F}_2(9)$, $\mathcal{F}_3(3)$ and $\mathcal{F}_4(3)$. 
Dashed lines represent negative edges, while continuous lines represent positive edges.}
\label{fig:validdrawinginducedsubgraphline}
\end{center}
\end{figure}
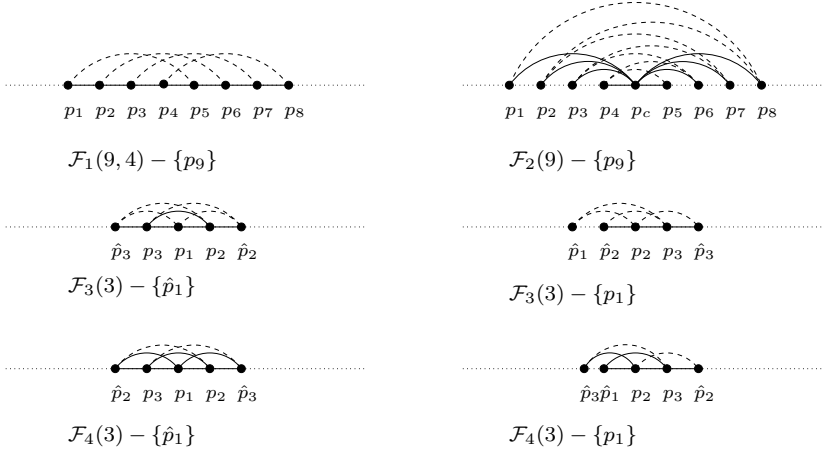

Now, we proceed with the signed graphs $\mathcal{F}_2(2n-1)$. In this case, when the central vertex is deleted, the induced subgraph that we obtain has no positive edge. Hence, any drawing is valid. On the other hand, 
when one of the vertices in the negative cycle is deleted, by symmetry, the graph we obtain is always the same, independent of the deleted vertex. Let us denote from $1$ to $2n - 1$ the vertices of the negative cycle of $\mathcal{F}_2(2n-1)$ clockwise according to the cycle. W.l.o.g. we delete vertex $2n-1$. Hence, the resulting graph has $2(n-1)$ vertices from the negative cycle plus the central vertex. A valid drawing for such a signed graph embeds the central vertex at the center, say at zero. Then, it places every even numbered vertex at the left of the central vertex and every odd numbered vertex at the right hand side of the central vertex. Vertices at the left and at the right hand side of the central vertex are placed in the first $n-1$ integer positions next to the central vertex. See Figure \ref{fig:validdrawinginducedsubgraphline} for an example of such a drawing applied to the induced subgraph of $\mathcal{F}_2(9)$.

Families $\mathcal{F}_3$ and $\mathcal{F}_4$ have a similar form. In the core, they have a cycle, plus external vertices connected to three consecutive vertices in the cycle. In order to show that these families do not have valid drawing in the line it was crucial the fact that, if we consider any embedding of   
a cycle of odd length in the line, there exists at least one vertex such that one neighbor is at its left side and the other neighbor is embedded at its right hand side. 
Now, we use the fact that there exists a embedding of a cycle of odd length in the line such that only one vertex has one neighbor at its left side and one neighbor at its right hand side. 
In order to obtain a valid drawing for induced subgraphs of $\mathcal{F}_3(2n-1)$ and  $\mathcal{F}_4(2n-1)$, we consider such an embedding. 

When one vertex of the cycle is deleted, w.l.o.g., we assume that the deleted vertex is the only vertex with neighbors at its left and right. When one of the external vertices is deleted, we assume that 
the deleted vertex corresponds to the only vertex with neighbors at its left and right in the cycle. Then, the external vertices can be arranged at the left or at the right hand side of the embedding of the cycle, 
according to the sign of the edges connecting those nodes with the cycle. 
%
\end{proof}

\section{Drawing signed graphs in the line}\label{sec:linecase}
In this section we focus in the special case $\ell = 1$, i.e., 
following problem: Given a signed graph $G$, we seek to determine
whether $G$ has a valid drawing in the Euclidean line.  We
prove that when the signed graph $G$ is complete, there exists a
polynomial time algorithm that determines whether a given signed graph
$G$ has a valid drawing in the line.  Moreover, in the case
that such a valid drawing exists, we give a polynomial time algorithm
that provides it.

We start by pointing out the fact that any drawing of a given signed
graph $G$ in the Euclidean line determines an ordering on the set of
vertices $V$. Let $D(G)$ be a drawing of $G$ in the Euclidean line.
Then, $D(G)$ is a set of values $\{u_{p_1},u_{p_2},\ldots,u_{p_n}\}$ in
$\mathbb{R}$, where $u_{p_i}$ determines the position in the line for
vertex $p_i$. The ordering in which we are interested follows: we say
that $p_i < p_j \iff u_{p_i} < u_{p_j}$.  Without loss of generality, we
assume that $u_{p_1} < u_{p_2} < \cdots < u_{p_n}$. Hence, the implicit ordering
on the set of vertices $V$ given by $D(G)$ is $p_1< p_2<\cdots <p_n$.
\begin{definition}
Given a signed graph $G=(V, E^+\bigcup E^-)$, and an ordering on the set of vertices $V$. We define $p_i$'s augmented incident vector 
$\overrightarrow{p_i} \in \{+1,0,-1\}^n$ as follow:
$$
\overrightarrow{p_i} = (p_{i1}, p_{i2}, \ldots, p_{in}), \quad \mbox{where} \quad p_{ij} =\left\{ \begin{array}{rcl}
-1 & \mbox{for}
& p_j\in P_i^- \\ 
1 & \mbox{for} & i=j \\
1 & \mbox{for} & p_j \in P_i^+\\
0 & \mbox{for} & p_ip_j\notin E.
\end{array}\right.
$$
Since the graph is undirected, it holds $p_{ij}=p_{ji}$. 
\end{definition}
The vector is called augmented since we define $p_{ii} = 1$ for all $i$. For simplicity we call this vector $p_i$'s vector. 

\begin{lemma}\label{lem:properdrawingtoorder}
Let $G=(V,E^+\bigcup E^-)$ be a signed graph. Let $p_1< p_2< \cdots< p_n$ be an ordering on the set of vertices $V$ given by a valid drawing
of $G$ in the Euclidean line. Then, for all $p_i$ it holds:
\begin{itemize}
\item[(i)] if $p_{ij}=-1$ and $p_j<p_i \Rightarrow \forall \: p_{j'}<p_j$, $p_{ij'} \in \{-1,0\}$, and 
\item[(ii)] if $p_{ij}=-1$ and $p_j>p_i \Rightarrow \forall \: p_{j'}>p_i$, $p_{ij'} \in \{-1,0\}$. 
\end{itemize}
\end{lemma}
\begin{proof}
The proof is by contradiction. Let us assume that, even if the ordering on the set of vertices $V$ comes from a valid 
drawing of $G$ in the line, 
condition \textit{(i)} does not hold, i.e.,
there exist $p_{j'}<p_j<p_i$ such that $p_{ij'}=1$ (equivalently, vertices $p_i$ and $p_{j'}$ are connected by a positive edge) 
and $p_{ij}=-1$ (equivalently, vertices $p_i$ and $p_j$ are connected by a negative edge). Since 
 these three vertices are ordered as $p_{j'}<p_j<p_i$, thus it holds: $d(p_i,p_{j'})>d(p_i,p_j)$.
Nevertheless, the previous condition is a contradiction because, the ordering comes from a 
valid drawing of $G$ in the line, thus it holds: $d(p_i,p_j)>d(p_i,p_{j'})$. 
The proof proceeds equivalently if we assume that condition \textit{(ii)} does not hold.
\end{proof}
 
We have seen that conditions $(i)$ and $(ii)$ are necessary for any valid drawing in the line. Now, we will see that, indeed, they are sufficient conditions for a valid drawing in the line. 
In order to do that. let us first introduce the following notation. 
Let $L^-(i)$ be the \emph{closest smaller negative neighbor} of vertex $p_i$ defined as follows: 
$$L^-(i) = \left\{p_j : p_j < p_i \: \wedge \:  (p_jp_i) \in E^- \: \wedge \: p_{ki} \in \{0,1\} \: \forall \: j <k < i \right\}.$$
Equivalently, let us denote $R^-(i)$ the \emph{closest 
bigger negative neighbor} of vertex $p_i$ defined as follows: 
$$
R^-(i) = \left\{p_j: p_i < p_j \: \wedge \: (p_ip_j) \in E^- \: \wedge \: p_{ik} \in \{0,1\} \: \forall \: i < k < j  \right\}.
$$
In order to complete this definitions, and to be consistent when these definitions are applied to vertices $p_1$ and $p_n$, we include two artificial vertices $p_0$ and $p_\infty$ 
such that $p_0<p_1$, $p_n<p_\infty$, and $p_{i0}=p_{i\infty}=-1$ for all vertex $p_i$. Then, 
$L^-(1)=\{p_0\}$ and $R^-(n)=\{p_\infty\}$.

We also define the \emph{farthest smaller positive neighbor} and the \emph{farthest bigger positive neighbor} 
of vertex $p_i$, denoted by $L^+(i)$ and $R^+(i)$ respectively, as follow: 
$$L^+(i) = \left\{p_j : p_j \leq p_i \: \wedge \:  (p_jp_i) \in E^+ \: \wedge \: p_{ki} \in \{-1,0\} \: \forall \: j <k < i \right\}.$$
and
$$
R^+(i) = \left\{p_j: p_i \leq p_j \: \wedge \: (p_ip_j) \in E^+ \: \wedge \: p_{ik} \in \{-1,0\} \: \forall \: i < k < j  \right\}.
$$
Note that each ob the above defined elements is a single vertex, hence we can treat them like that. 
Furthermore, in the special case 
when an ordering on the set of vertices $V$ satisfies conditions $(i)$ and $(ii)$, 
it holds $$L^-(i) < L^+(i) \leq p_i \leq R^+ (i) < R^-(i).$$ That is an important characteristic in what follows.

\begin{lemma}\label{lem:ordertoproperdrawing}
Let $G=(V,E^+\bigcup E^-)$ be a signed graph. If there exists an ordering on the set of vertices $V$ such that
for all $p_i$,  it holdst: 
\begin{itemize}
\item[(i)] if $p_{ij}=-1$ and $p_j<p_i \Rightarrow \forall \: p_{j'}<p_j$, $p_{ij'} \in \{-1,0\}$, and 
\item[(ii)] if $p_{ij}=-1$ and $p_j>p_i \Rightarrow \forall \: p_{j'}>p_i$, $p_{ij'} \in \{-1,0\}$. 
\end{itemize}
then, there exists a valid drawing of $G$ in the line. 
\end{lemma}
\begin{proof}
Let $G$ be a signed graph, and $\mathcal{O} = p_1<p_2<p_3<\cdots<p_n$ be an ordering on the set of vertices $V$ that satisfies condition $(i)$ and $(ii)$. Let us denote $u_{p_i}$ the position of vertex $p_i$. 
The proof is constructive. We assign real values to every position $u_{p_i}$.
The construction maintains ordering $\mathcal{O}$, i.e., two vertices $p_i$ and $p_j$ in $V$
are placed in the line at $u_{p_i}$ and $u_{p_j}$ respectively, such that if $p_i<p_j$ then $u_{p_i} <u_{p_j}$.

We first point out the fact that, in order to obtain a valid drawing, it is enough to guarantee for all vertex $p_i$ the next four conditions: 
\begin{eqnarray*}
&&d(L^+(i),p_i) < d(L^-(i),p_i)\: \& \:
d(p_i,R^+(i)) < d(p_i,R^-(i)), \\ 
&&d(L^+(i),p_i) < d(p_i,R^-(i)) \: \&\:
d(p_i,R^+(i)) < d(L^+(i),p_i).
\end{eqnarray*}

Since the construction follows the ordering $\mathcal{O}$, and the ordering $\mathcal{O}$ satisfies 
conditions $(i)$ and $(ii)$, it holds $L^-(i) < L^+(i) 	\leq p_i \leq  R^+ (i) < R^-(i)$, for all vertex $p_i$. 
Thus, it also holds $u_{L^-(i)} < u_{L^+(i)} \leq u_{p_i} \leq u_{R^+ (i)} < u_{R^-(i)}$.
Therefore, since the ordering $\mathcal{O}$ satisfies 
conditions $(i)$ and $(ii)$, for every vertex it holds that every smaller (resp. bigger) positive neighbor is placed 
closer than every smaller (resp. bigger) negative neighbor, or equivalently the two first conditions previously presented are fulfilled $d(L^+(i),p_i) < d(L^-(i),p_i)\: \& \:
d(p_i,R^+(i)) < d(p_i,R^-(i))$. 


Hence, in order to construct a valid drawing, we  need to guarantee for all vertex $p_i$ the two second conditions 
$d(L^+(i),p_i) < d(p_i,R^-(i)) \: \&\:
d(p_i,R^+(i)) < d(L^+(i),p_i)$.
In oder words, we need
to guarantee that the drawing satisfies 
$ u_{R^+ (i)} -u_{p_i} < u_{p_i} - u_{L^-(i)} \mbox{ and } u_{p_i} - u_{L^+(i)} < u_{R^-(i)} - u_{p_i}$
for every vertex $p_i$. An equivalent way to present these two conditions follows: 
\begin{equation}\label{eq:condition}
u_{R^+(i)}+ u_{L^-(i)} <2u_{p_i}<u_{R^-(i)} +u_{L^+(i)}
\end{equation}
Now, we give a construction that provides such guarantee. 


The construction is sequential, it sets arbitrarily $u_{p_0} = 0$. Then, it continues in order giving values to  vertices $u_{p_1},u_{p_2}, \ldots, u_{p_n}$ according condition (\ref{eq:condition}). 
Therefore, when $u_{p_i}$ is determined, it imposes a condition on some $u_{p_j}$ for $p_j$ strictly bigger than $p_i$.  
Since, by construction the artificially included vertex $p_0$ is connected negatively with every other vertex, then $R^+(0) = L^+(0) = \{\emptyset\}$ and $R^-(0) = \{p_n\}$. On the other hand, 
since vertex $p_0$ is the leftmost vertex in the ordering, then $L^-(0) = \{\emptyset\}$. Hence, when condition (\ref{eq:condition}) is applied to $p_0$, it says $0 < u_{p_n}$.

Again, the construction arbitrarily sets $u_{p_1}=1$. By definition of $p_0$, $L^-(i)$, and$L^+(i)$, we have that $L^-(1) = \{p_0\}$, $L^+(0)= \{p_1\}$. Hence, when condition (\ref{eq:condition}) is applied to $p_1$, it imposes $u_{R^+(1)} < 2 < u_{R^-}(1) - 1$. Hence, when the construction determines values for  $u_{R^+(1)}$ and $u_{R^-}(1)$ it has to take in consideration this restriction. The construction follows sequentially according to the ordering $\mathcal{O}$ taking in consideration all the restriction that each assignment produces for future assignments. 


The proof finishes showing that these conditions are satisfiable. 
We prove this last point by contradiction. Assume there are two conditions that contradict each other, i.e., 
for some $p_j<p_i$ such that $u_{R^+(i)}=u_{R^-(j)}$, the construction has to guarantee that $u_{R^+(i)} <2u_{p_i}- u_{L^-(i)}$
and $2u_{p_j}-u_{L^+(j)}<u_{R^-(i)}$, but the assignment produces $2u_{p_i}- u_{L^-(i)} <2u_{p_j}-u_{L^+(j)}$.

If we have $2u_{p_i}- u_{L^-(i)} <2u_{p_j}-u_{L^+(j)}$, then also we have $2u_{p_i} - 2u_{p_j} < u_{L^-(i)}-u_{L^+(j)}$. On the other hand, 
the assignment for $u_{p_i}$ and $u_{p_j}$ followed condition (\ref{eq:condition}), hence $2u_{p_i} - 2u_{p_j} > u_{R^+(i)}+ u_{L^-(i)} - u_{R^-(j)} -u_{L^+(j)}$.
Since $u_{R^+(i)}=u_{R^-(j)}$, the previous equation is equivalent to $2u_{p_i} - 2u_{p_j} > u_{L^-(i)} -u_{L^+(j)}$, which generates a contradiction. 
Then, when the construction sets value for $u_{p_i}$, it can satisfy all the conditions produced for $u_{p_i}$ until that point. 
Hence, it produces a valid drawing in the Euclidean line and the Lemma is proved. 
\end{proof}

Until this point, we have characterized the property of having a valid
drawing by the existence of an ordering on the set of vertices such
that conditions $(i)$ and $(ii)$ are satisfied for every vertex.
Therefore, determining whether a given a signed graph $G$ has a valid
drawing in the Euclidean line is equivalent to determine whether the set of vertices of $G$ has
an ordering such that conditions $(i)$ and
$(ii)$ are satisfied for every vertex. In the following, we focus on
that task. From now on, we say that a signed graph is \emph{complete}
if its underling graph is a clique.  Also, we define $G^+$ to be the
\emph{positive graph} of a given signed graph $G$ as the subgraph of
$G$ composed by its positive edges, i.e., $G^+=(V,E^+)$. The positive
graph of a given signed graph is not considered a signed graph.

\begin{lemma}\label{lem:completesignedgraph}
  Let $G$ be a complete signed graph, and $G^+$ be its positive
  graph. If there exists an ordering $\mathcal{O}$ on the set of
  vertices $V$ such that conditions $(i)$ and $(ii)$ are satisfied for
  every vertex, then $G^+$ is chordal.
\end{lemma}  
\begin{proof}
  Let us remind you that a \emph{chordal graph} is a graph, not
  necessarily signed, where every induced cycle on four or more
  vertices has a chord.  Let $p_1,p_2,\ldots,p_l$ be a set of vertices
  that form a cycle in $G^+$. Without loss of generality, let us
  assume that $p_1<p_2<\cdots<p_l$ following ordering $\mathcal{O}$.
  Since $G$ is a complete graph, conditions $(i)$ and $(ii)$ are satisfied for every vertex and
  $p_{11}=p_{1l}=1$ (because $p_1$ and $p_l$ are in a cycle of positive
  edges), then $p_{1i}=1$ for all $1\leq i\leq l$. Therefore,
  particularly there exists a chord always when $l \geq 4$. Hence, $G^+$ is chordal.
\end{proof}

An ordering $p_1<p_2<\cdots<p_n$ of vertices is a \emph{perfect
  elimination ordering} of graph $G$ if the neighborhood of each
vertex $p_i$ forms a clique in the graph induced by vertices
$p_i,\ldots,p_n$.  It is known that a graph is chordal if and only if
there exists \emph{perfect elimination ordering} on its set of
vertices. (See e.g. \cite{DBLP:journals/tcs/HabibMPV00}).
\begin{lemma}\label{lem:completesignedgraphwithvaliddrawing}
  Let $G$ be a complete signed graph with a valid drawing in the
  Euclidean line, and with more than four vertices. Let $G^+$ be $G$'s
  positive graph.  If $G^+$ is connected then every perfect
  elimination ordering on the set of vertices satisfies
  conditions $(i)$ and $(ii)$ for every vertex.
\end{lemma}
\begin{proof}
Consider a complete signed graph $G$ with a valid drawing in the Euclidean line, 
and with more than four vertices. Assume that $G^+$, $G$'s positive graph, is connected. 
The proof is by contradiction. Hence, let us assume that $\mathcal{O}$ is a perfect elimination ordering for $G^+$, but either condition $(i)$ is 
not satisfied or condition $(ii)$ is not satisfied. Let us assume that condition $(i)$ is not satisfied. Then, there exist vertices $p_i<p_j<p_l$ such that 
$p_{lj}=-1$ and $p_{li}=1$. Since the ordering is a perfect elimination ordering, then $p_{ij}=-1$, otherwise $p_{lj}=1$. Now, since there exists a fourth vertex
$p_e$ and $G^+$ is connected, then $p_{ej}=1$ and either $p_{ei}=1$ or $p_{el}=1$. But does not matter the position in the ordering for $p_e$, since the ordering is a perfect elimination ordering, 
then either $p_{ij}=1$ when $p_{ei}=1$, or $p_{jl}=1$ when $p_{el}=1$. In both cases there is a contradiction. The proof for the case
when condition $(ii)$ is not satisfied follows equivalently. Hence a perfect elimination ordering has to be such that conditions $(i)$ and $(ii)$ are satisfied for every vertex.
\end{proof}

In \cite{DBLP:journals/tcs/HabibMPV00}, the authors present an algorithm that decides whether a graph is chordal, and computes a perfect elimination 
ordering in case it exists, in time $O(n+m)$, where $n$ is the number of vertices and $m$ is the number of edges. Therefore, using that algorithm and the previous results stated in this section, 
we can decide whether a given complete signed graph has a valid drawing and compute such a drawing in polynomial time.  
\begin{theorem}\label{thm:final}
Given a complete signed graph $G$, the question of whether it has a valid drawing in the Euclidean line can be computed in time $O(n^2)$. Furthermore, if $G$ has a valid drawing
in the Euclidean line, computing such a drawing can be done in time $O(n^2)$. 
\end{theorem}
\begin{proof}
Let $G$ be a complete signed graph, and $G^+$ be its positive graph. Let $G^+_1, G^+_2, \ldots, G^+_k$ be the connected components of $G^+$. Then, using algorithm presented in \cite{DBLP:journals/tcs/HabibMPV00}, 
decide for each $G^+_i$ whether $G^+_i$ is a chordal graph and compute a perfect elimination ordering in case it exists. By Lemma \ref{lem:completesignedgraph}, if there exists $G^+_i$ 
such that it is not chordal, then $G$ has not a valid drawing. If all $G^+_1, G^+_2, \ldots, G^+_k$ are chordal, then by Lemma \ref{lem:completesignedgraphwithvaliddrawing} a perfect
elimination ordering for each of them is such that conditions $(i)$ and $(ii)$ are satisfied in its induced signed subgraph. We produce 
an ordering on the set of vertices $V$ by a concatenation of the orderings for $G^+_i$. Hence, given two vertices $p_i$ and $p_j$ if they form part of different connected components, let us say
$p_i\in G^+_l$ and $p_j\in G^+_q$, then $p_i<p_j \iff l<q$. On the other hand, if $p_i$ and $p_j$ form part of the same connected component, then the perfect elimination ordering decides the order 
between them. Such an ordering makes satisfiable conditions $(i)$ and $(ii)$ for every vertex. The last statement is proved by contradiction. Let us assume that condition $(i)$ is not satisfied.
Then, there exists a vertex $p_i<p_j<p_l$ such that $p_{il}=1$ and $p_{jl}=-1$. Since, $p_{il}=1$, then vertices $p_i$ and $p_l$ form part of the same connected component. Due to the concatenation of the orderings, 
then $p_j$ is part of the same connected component. But then there is a contradiction, because in a connected component vertices follow a perfect elimination ordering, and it satisfies condition $(i)$ and $(ii)$. 
The proof follows equivalently when condition $(ii)$ is not satisfied. Therefore, the concatenation of the orderings satisfies conditions $(i)$ and $(ii)$ for every vertex.
Finally, the construction presented in the Proof of Lemma \ref{lem:ordertoproperdrawing} produces a valid drawing for $G$. Since, the perfect elimination ordering can be computed in time $O(n+m)$, the verification that such ordering satisfies conditions $(i)$ and $(ii)$ can be done in time $O(n^2)$ and the drawing of Lemma \ref{lem:ordertoproperdrawing} is computed in time $O(n^2)$, then the decision problem is answered in time $O(n^2)$ and, in case it is applicable, the construction of the valid drawing is done in time $O(n^2)$.
\end{proof}

%
%
%
%
\section{Conclusions and future work}\label{sec:futurework}
An extended abstract of this work was published in \cite{DBLP:conf/mfcs/KermarrecT11}. Subsequently, M. Cygan et al. proved in \cite{DBLP:conf/mfcs/CyganPPW12} that the general version of the decision problem studied in this document (i.e., the question answered in Theorem \ref{thm:final} but not restricted to complete graphs) is NP-Complete. 

Nonetheless, this work still leaves open many interesting research directions and this is precisely one of the strength of this piece of work. 
%
 Is it easier in any metric space find a valid draiwng when 
the given signed graph is complete as well as it is in the case of the Euclidean line? 
%
Extensions to  the problem might go along different directions. For instance, the metric space can be different, it might not be Euclidean. A more general question is to study the impact of the metric space, for instance, 
on the forbidden patterns. On the other hand, the value assignment to the edges can range in a larger set of values rather than being binary. The last proposed extension is interesting since it enables to model recommendation systems with ratings in a more precise way. 

 
\bibliographystyle{plain}
\bibliography{signed}

\begin{thebibliography}{10}

\bibitem{PhysRevE.72.036121}
T.~Antal, P.~L. Krapivsky, and S.~Redner.
\newblock Dynamics of social balance on networks.
\newblock {\em Phys. Rev. E}, 72(3):036121, Sep 2005.

\bibitem{DBLP:journals/ml/BansalBC04}
Nikhil Bansal, Avrim Blum, and Shuchi Chawla.
\newblock Correlation clustering.
\newblock {\em Machine Learning}, 56(1-3):89--113, 2004.

\bibitem{DBLP:journals/tvcg/BrandesFL06}
Ulrik Brandes, Daniel Fleischer, and J{\"u}rgen Lerner.
\newblock Summarizing dynamic bipolar conflict structures.
\newblock {\em IEEE Trans. Vis. Comput. Graph.}, 12(6):1486--1499, 2006.

\bibitem{Cartwright1956277}
Dorwin Cartwright and Frank Harary.
\newblock Structural balance: a generalization of heider's theory.
\newblock {\em Psychological Review}, 63(5):277 -- 293, 1956.

\bibitem{DBLP:conf/mfcs/CyganPPW12}
Marek Cygan, Marcin Pilipczuk, Michal Pilipczuk, and Jakub~Onufry Wojtaszczyk.
\newblock Sitting closer to friends than enemies, revisited.
\newblock In Branislav Rovan, Vladimiro Sassone, and Peter Widmayer, editors,
  {\em MFCS}, volume 7464 of {\em Lecture Notes in Computer Science}, pages
  296--307. Springer, 2012.

\bibitem{davis67}
James~A. Davis.
\newblock Clustering and structural balance in graphs.
\newblock {\em Human Relations}, 20(2):181 -- 187, 1967.

\bibitem{DBLP:journals/tcs/HabibMPV00}
Michel Habib, Ross~M. McConnell, Christophe Paul, and Laurent Viennot.
\newblock Lex-bfs and partition refinement, with applications to transitive
  orientation, interval graph recognition and consecutive ones testing.
\newblock {\em Theor. Comput. Sci.}, 234(1-2):59--84, 2000.

\bibitem{harary53}
Frank Harary.
\newblock On the notion of balance of a signed graph.
\newblock {\em Michigan Mathematical Journal}, 2(2):143 -- 146, 1953.

\bibitem{DBLP:journals/mss/HararyK80}
Frank Harary and Jerald~A. Kabell.
\newblock A simple algorithm to detect balance in signed graphs.
\newblock {\em Mathematical Social Sciences}, 1(1):131--136, 1980.

\bibitem{harary81}
Frank Harary and Jerald~A. Kabell.
\newblock Counting balanced signed graphs using marked graphs.
\newblock {\em Proceedings of the Edinburgh Mathematical Society},
  24(2):99--104, 1981.

\bibitem{springerlink:10.1007/BF02476926}
Frank Harary and Edgar Palmer.
\newblock On the number of balanced signed graphs.
\newblock {\em Bulletin of Mathematical Biology}, 29:759--765, 1967.

\bibitem{DBLP:conf/mfcs/KermarrecT11}
Anne-Marie Kermarrec and Christopher Thraves.
\newblock Can everybody sit closer to their friends than their enemies?
\newblock In Filip Murlak and Piotr Sankowski, editors, {\em MFCS}, volume 6907
  of {\em Lecture Notes in Computer Science}, pages 388--399. Springer, 2011.

\bibitem{DBLP:conf/sdm/KunegisSLLLA10}
J{\'e}r{\^o}me Kunegis, Stephan Schmidt, Andreas Lommatzsch, J{\"u}rgen Lerner,
  Ernesto William~De Luca, and Sahin Albayrak.
\newblock Spectral analysis of signed graphs for clustering, prediction and
  visualization.
\newblock In {\em SDM}, pages 559--570. SIAM, 2010.

\bibitem{DBLP:conf/cse/LauterbachTSA09}
Debra Lauterbach, Hung Truong, Tanuj Shah, and Lada~A. Adamic.
\newblock Surfing a web of trust: Reputation and reciprocity on
  couchsurfing.com.
\newblock In {\em CSE (4)}, pages 346--353. IEEE Computer Society, 2009.

\bibitem{DBLP:conf/icwsm/LeskovecHK10}
Jure Leskovec, Daniel~P. Huttenlocher, and Jon~M. Kleinberg.
\newblock Governance in social media: A case study of the wikipedia promotion
  process.
\newblock In William~W. Cohen and Samuel Gosling, editors, {\em ICWSM}. The
  AAAI Press, 2010.

\bibitem{DBLP:conf/www/LeskovecHK10}
Jure Leskovec, Daniel~P. Huttenlocher, and Jon~M. Kleinberg.
\newblock Predicting positive and negative links in online social networks.
\newblock In Michael Rappa, Paul Jones, Juliana Freire, and Soumen Chakrabarti,
  editors, {\em WWW}, pages 641--650. ACM, 2010.

\bibitem{DBLP:conf/chi/LeskovecHK10}
Jure Leskovec, Daniel~P. Huttenlocher, and Jon~M. Kleinberg.
\newblock Signed networks in social media.
\newblock In Elizabeth~D. Mynatt, Don Schoner, Geraldine Fitzpatrick, Scott~E.
  Hudson, W.~Keith Edwards, and Tom Rodden, editors, {\em CHI}, pages
  1361--1370. ACM, 2010.

\bibitem{minsa}
Eduardo~G. Pardo, Mauricio Soto, and Christopher Thraves.
\newblock Embedding signed graphs in the line.
\newblock {\em Journal of Combinatorial Optimization}, pages 1--21, 2013.

\bibitem{szell10}
Michael Szell, Renaud Lambiotte, and Stefan Thurner.
\newblock Multirelational organization of large-scale social networks in an
  online world.
\newblock {\em PNAS}, 107(31):13636--13641, 2010.

\end{thebibliography}

\end{document}